\documentclass{LMCS}

\def\doi{9(2:07)2013}
\lmcsheading%
{\doi}
{1--34}
{}
{}
{Mar.~30, 2012}
{Jun.~21, 2013}
{}

\usepackage{amssymb}
\usepackage{amsmath}
\usepackage{graphicx}
\usepackage{gastex}
\usepackage{ifpdf}
\usepackage{color}
\usepackage{wrapfig}
\usepackage{gastex}
\usepackage{enumerate,hyperref}

\def\phi{\varphi}
\def\epsilon{\varepsilon}
\DeclareMathOperator{\interleave}{||}
\DeclareMathOperator{\U}{\text{\sf U}}
\DeclareMathOperator{\X}{\text{\sf X}}
\newtheorem{counter}{Counterexample}

\def\tilde{\widetilde}

\newcommand{\TRANA}[3]{#1\xrightarrow[]{#2}#3}
\newcommand{\TRAN}[2]{#1\rightarrow #2}
\newcommand{\TRANP}[2]{#1\rightarrow_{{\sf P}}#2}
\newcommand{\BSP}{\ensuremath{\sim_{{\sf P}}}}
\newcommand{\nBSP}{\ensuremath{\nsim_{{\sf P}}}}

\newcommand{\Si}{\prec}

\newcommand{\AP}{\mathit{AP}}
\newcommand{\PA}{{\sf PA}}
\newcommand{\SP}{\ensuremath{\prec_{{\sf P}}}}

\newcommand{\DTMC}{{\sf DTMC}}

\newcommand{\boundTRAN}[2]{\overset{#1,#2}{\Longrightarrow}}

\newcommand{\WBS}{\approx}

\newcommand{\WSi}{\precapprox}

\newcommand{\PCTL}{{\sf PCTL}}
\newcommand{\PCTLS}{{\sf PCTL}^{*}}

\newcommand{\CTL}{{\sf CTL}}

\newcommand{\iBS}[1]{\sim_{#1}}

\newcommand{\infBS}{\sim}
\newcommand{\iBSB}[1]{\sim_{#1}^{{\sf b}}}

\newcommand{\infBSB}{\sim^{{\sf b}}}
\newcommand{\EPCTL}{\sim_{\PCTL}}
\newcommand{\EPCTLM}{\sim_{\PCTL^{-}}}
\newcommand{\iEPCTLM}[1]{\sim_{\PCTL^{-}_{#1}}}
\newcommand{\inEPCTLM}[1]{\nsim_{\PCTL^{-}_{#1}}}
\newcommand{\EPCTLS}{\sim_{\PCTL^{*}}}
\newcommand{\nEPCTLS}{\nsim_{\PCTL^{*}}}
\newcommand{\EPCTLSM}{\sim_{\PCTL^{*-}}}
\newcommand{\iEPCTLSM}[1]{\sim_{\PCTL_{#1}^{*-}}}
\newcommand{\nEPCTL}{\nsim_{\PCTL}}

\newcommand{\iSi}[1]{\prec_{#1}}

\newcommand{\iBSi}[1]{\prec_{#1}^{{\sf b}}}

\newcommand{\SEPCTL}{\prec_{\PCTL}}
\newcommand{\SEPCTLM}{\prec_{\PCTL^{-}}}
\newcommand{\iSEPCTLM}[1]{\prec_{\PCTL^{-}_{#1}}}
\newcommand{\SEPCTLS}{\prec_{\PCTL^{*}}}

\newcommand{\SEPCTLSM}{\prec_{\PCTL^{*-}}}
\newcommand{\iSEPCTLSM}[1]{\prec_{\PCTL_{#1}^{*-}}}

\newcommand{\PAR}[2]{#1\interleave#2}

\newcommand{\bBSP}{\simeq_{{\sf P}}}

\newcommand{\bSiP}{\precapprox_{{\sf P}}}

\newcommand{\MC}[1]{\mathcal{#1}}
\newcommand{\MI}[1]{\mathit{#1}}
\newcommand{\DSI}[1][\MC{R}]{\sqsubseteq_{#1}}

\newcommand{\DIRAC}[1]{\mathcal{D}_{#1}}

\newcommand{\DEPTH}{\mathit{Depth}}
\newcommand{\EPCTLWN}{\sim_{\PCTL_{\backslash\X}}}
\newcommand{\EPCTLSWN}{\sim_{\PCTL^{*}_{\backslash\X}}}

\newcommand{\nEPCTLSWN}{\nsim_{\PCTL^{*}_{\backslash\X}}}

\newcommand{\SEPCTLWN}{\precapprox_{\PCTL_{\backslash\X}}}
\newcommand{\SEPCTLSWN}{\precapprox_{\PCTL^{*}_{\backslash\X}}}

\newcommand{\WBSB}{\approx^{{\sf b}}}

\newcommand{\WBSi}{\precapprox^{{\sf b}}}

\newcommand{\bTRAN}[2]{#1\Rightarrow^{\MC{R}}#2}
\newcommand{\bTRANP}[2]{#1\Rightarrow^{\MC{R}}_{\text{P}}#2}

\newcommand{\MEASURE}{\mathit{Prob}}
\newcommand{\MEASUREONE}{\mathit{PreCap}}
\newcommand{\MEASURETWO}{\mathit{PostCap}}

\newcommand{\SUP}{\mathit{Sup}}
\newcommand{\ABS}[1]{|#1|}
\newcommand{\SUPP}{\mathit{Supp}}

\newcommand{\DOWNWARD}[2]{#1^{\downarrow}#2}

\begin{document}
\title[Bisimulations Meet PCTL Equivalences for Probabilistic Automata]{Bisimulations Meet PCTL Equivalences for Probabilistic Automata\rsuper*}

\author[L.~Song]{Lei Song\rsuper a}	
\address{{\lsuper a}Max-Planck-Institut f\"{u}r Informatik, and
Saarland University -- Computer Science, Germany}	 
\email{song@cs.uni-saarland.de}  

\author[L.~Zhang]{Lijun Zhang\rsuper b}	
\address{{\lsuper b}State Key Laboratory of Computer Science, Institute of Software, Chinese Academy of Sciences,
DTU Informatics,  Technical University of Denmark, and
Saarland University -- Computer Science, Germany}	 
\email{zhang@imm.dtu.dk and zhanglj@ios.ac.cn (corresponding author)}  

\author[J. C. Godskesen]{Jens Chr. Godskesen\rsuper c}	
\address{{\lsuper c}IT University of Copenhagen, Denmark}	 
\email{jcg@itu.dk}  

\author[F. Nielson]{Flemming Nielson\rsuper d}	
\address{{\lsuper d}DTU Compute,  Technical University of Denmark}	
\email{fnie@dtu.dk}  

\ACMCCS{[{\bf  Mathematics of computing}]: Probability and statistics---Stochastic processes---Markov processes;  [{\bf Theory of computation}]: Logic---Modal and temporal logics;  Semantics and reasoning---Program reasoning---Program verification; [{\bf General and reference}]: Cross-computing tools and techniques---Performance; Cross-computing tools and techniques---Verification}

\titlecomment{{\lsuper*}An extended abstract of the paper has appeared in \cite{SongZG2011}}

\begin{abstract}
  Probabilistic automata ($\PA$s) have been successfully applied in
  formal verification of concurrent and stochastic systems. Efficient
  model checking algorithms have been studied, where the most often
  used logics for expressing properties are based on probabilistic computation tree logic ($\PCTL$) and its
  extension $\PCTL^{*}$. Various behavioral equivalences are proposed,
  as a powerful tool for abstraction and compositional minimization
  for $\PA$s. Unfortunately, the equivalences are
  well-known to be sound, but not complete with respect to the logical
  equivalences induced by $\PCTL$ or $\PCTL^{*}$.  The desire of a
  both sound and complete behavioral equivalence has been pointed
  out by Segala in~\cite{Segala-thesis}, but remains open throughout
  the years.  In this paper we introduce novel notions
  of strong bisimulation relations, which characterize $\PCTL$ and
  $\PCTL^{*}$ exactly. We extend weak bisimulations that characterize $\PCTL$
  and $\PCTL^{*}$ without next operator, respectively.  Further, we
  also extend the framework to simulation preorders. Thus, our paper bridges
  the gap between logical and behavioral equivalences and preorders in this setting.
\end{abstract}
\maketitle
\section{Introduction}
Probabilistic automata ($\PA$s)~\cite{SegalaL95} have been
successfully applied in formal verification of concurrent and
stochastic systems. Efficient model checking algorithms have been
studied, where properties are mostly expressed by the probabilistic
computation tree logic ($\PCTL$)~\cite{hansson1994logic} and its
extension $\PCTL^{*}$~\cite{Aziz1995UWT} for Markov chains, and later
extended in~\cite{bianco1995model} for Markov decision processes.

To combat the infamous state space problem in model checking, various
behavioral equivalences, including strong and weak bisimulations, are
proposed for stochastic models including
$\PA$s~\cite{LarsenS89,larsen1991bisimulation,Segala-thesis,baier2003comparative,SegalaL95}. Indeed,
they turn out to be a powerful tool for abstraction for $\PA$s, since
bisimilar states imply that they satisfy exactly the same $\PCTL$ and
$\PCTL^{*}$ formulas, thus can be grouped together,
allowing one to construct smaller quotient automata before analyzing
the models.  In practice, bisimulation based minimization is
extensively studied in the literatures to leverage the state space
explosion, for instance see~\cite{CattaniS02,BaierEM00,KatoenKZJ07}.
Moreover, the nice compositional theory for $\PA$s is exploited for
compositional minimization~\cite{BoudaliCS09}, namely minimizing the
automata before composing the components together.

An interesting question is whether the reverse holds as well, namely whether logical equivalences imply bisimulation equivalences?
For Markov chains, i.e.,
$\PA$s without non-deterministic choices, the answer is affirmative~\cite{Aziz1995UWT,BaierKHW05}.
Unfortunately, the completeness does not hold in general, namely $\PCTL$
equivalence is strictly coarser than bisimulation and its variant
\emph{probabilistic bisimulation}~\cite{SegalaL95} for $\PA$s.

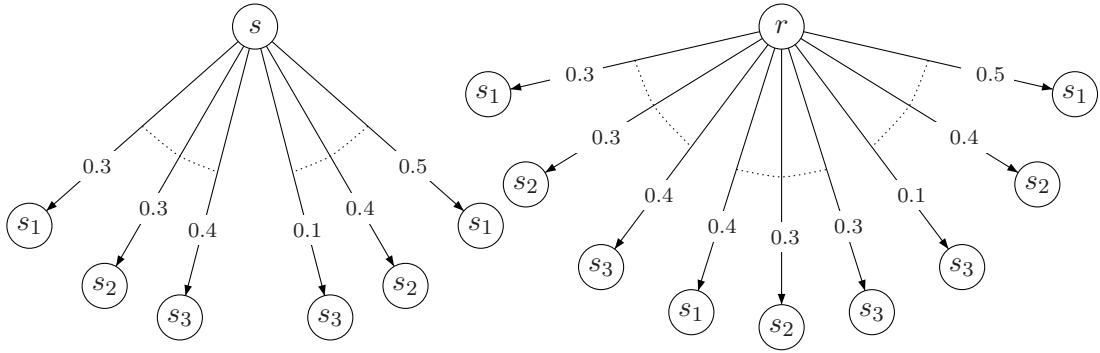
\begin{figure}[!t]
  \begin{center}
    \begin{picture}(80, 40)(0,20)
    \gasset{Nadjust=n,Nw=6, Nh=6,Nmr=3}
    \node(S)(0,60){$s$}
    \node(Saa)(-30,33.5){$s_1$}
    \node(Sba)(-20,25.5){$s_2$}
    \node(Sca)(-10,21.3){$s_3$}
    \node(Sab)(30,33.5){$s_1$}
    \node(Sbb)(20,25.5){$s_2$}
    \node(Scb)(10,21.3){$s_3$}
    \node(R)(70,60){$r$}
    \node(Raa)(31,51){$s_1$}
    \node(Rba)(36,39){$s_2$}
    \node(Rca)(46,28){$s_3$}
    \node(Rab)(58,22){$s_1$}
    \node(Rbb)(70,20){$s_2$}
    \node(Rcb)(82,22){$s_3$}
    \node(Rac)(109,51){$s_1$}
    \node(Rbc)(104,39){$s_2$}
    \node(Rcc)(94,28){$s_3$}
    \drawcurve[AHnb=0,ATnb=0,dash={0.2 0.5}0](-15,46.75)(-10,42.75)(-5,40.65)
    \drawcurve[AHnb=0,ATnb=0,dash={0.2 0.5}0](15,46.75)(10,42.75)(5,40.65)
    \drawcurve[AHnb=0,ATnb=0,dash={0.2 0.5}0](50.5,55.5)(53,49.5)(58,44)
    \drawcurve[AHnb=0,ATnb=0,dash={0.2 0.5}0](64,41)(70,40)(76,41)
    \drawcurve[AHnb=0,ATnb=0,dash={0.2 0.5}0](89.5,55.5)(87,49.5)(82,44)
    \gasset{ELdistC=y,ELdist=0}
    \drawedge[ELside=r,ELpos=70](S,Saa){{\scriptsize\colorbox{white} {0.3}}}
    \drawedge[ELside=r,ELpos=70](S,Sba){{\scriptsize\colorbox{white}{ 0.3}}}
    \drawedge[ELside=r,ELpos=70](S,Sca){{\scriptsize \colorbox{white}{0.4}}}
    \drawedge[ELpos=70](S,Sab){{\scriptsize\colorbox{white} {0.5}}}
    \drawedge[ELpos=70](S,Sbb){{\scriptsize\colorbox{white} {0.4}}}
    \drawedge[ELpos=70](S,Scb){{\scriptsize\colorbox{white} {0.1}}}
    \drawedge[ELside=r,ELpos=70](R,Raa){{\scriptsize\colorbox{white}{ 0.3}}}
    \drawedge[ELside=r,ELpos=70](R,Rba){{\scriptsize\colorbox{white}{ 0.3}}}
    \drawedge[ELside=r,ELpos=70](R,Rca){{\scriptsize\colorbox{white}{ 0.4}}}
    \drawedge[ELside=r,ELpos=70](R,Rab){{\scriptsize\colorbox{white}{ 0.4}}}
    \drawedge[ELside=r,ELpos=70](R,Rbb){{\scriptsize\colorbox{white}{ 0.3}}}
    \drawedge[ELside=r,ELpos=70](R,Rcb){{\scriptsize\colorbox{white}{ 0.3}}}
    \drawedge[ELpos=70](R,Rac){{\scriptsize\colorbox{white}{ 0.5}}}
    \drawedge[ELpos=70](R,Rbc){{\scriptsize\colorbox{white}{ 0.4}}}
    \drawedge[ELpos=70](R,Rcc){{\scriptsize\colorbox{white}{ 0.1}}}
    \end{picture}
    \end{center}
  \caption{\label{fig:counterexample}Counterexample of strong probabilistic bisimulation.}
\end{figure}

The main reason for the gap can be illustrated by the following
example. Consider the $\PA$ in Fig.~\ref{fig:counterexample}
assuming that $s_1,s_2,s_3$ are three absorbing states with different
state properties. It is easy to see that  $s$ and $r$ are $\PCTL$ equivalent:
the additional middle transition out of $r$ does not change the extreme probabilities,
the intervals of probabilities in which the three observing states can be reached are not changed.
However existing bisimulations differentiate $s$ and $r$,
mainly because the middle transition out of $r$ cannot be matched by
any transition (or combined transition) of $s$. Bisimulation requires
that the complete distribution of a transition must be matched, which is in this case too
strong, as it differentiates states satisfying the same $\PCTL$
formulas.

For $\PA$s, the desire of a both sound and complete behavioral
equivalence has been pointed out in~\cite{Segala-thesis} (see section
13.2.7), but remains open throughout the years.
Such a sound and complete relation is not only of theoretical interests: in practice
it would allow us to construct the minimal quotient automata for
checking $\PCTL$ and $\PCTL^{*}$ formulas, which are arguably the most often used
logics for specifying properties over $\PA$s.
 In this paper we bridge this gap by introducing novel
notions of behavioral equivalences which characterize (both soundly
and completely) $\PCTL$, $\PCTL^{*}$ and their sub logics.
Summarizing, our contributions are:
\begin{iteMize}{$\bullet$}
\item A new  bisimulation characterizing
$\PCTL^{*}$ soundly and completely. The bisimulation arises from a converging sequence of equivalence relations, each of which characterizes bounded $\PCTL^{*}$.
\item Branching bisimulations which correspond to $\PCTL$ and bounded
  $\PCTL$ equivalences.
\item We then extend our definitions to weak bisimulations, which characterize sub logics of $\PCTL$ and $\PCTL^{*}$ with only unbounded path formulas.
\item Further, we extend the
  framework to simulations as well as their characterizations,
extend the results to countable states, and discuss the coarsest congruent bisimulation  and simulation relations.
\end{iteMize}

\paragraph{Organization of the paper.}
Section \ref{sec:pre} introduces some notations. In Section \ref{sec:pa} we recall definitions of probabilistic automata and bisimulation relations by Segala~\cite{Segala-thesis}. We also recall the logic $\PCTL^{*}$ and its sub logics. Section \ref{sec:strong} introduces the novel strong and strong branching
bisimulations, and proves that they agree with $\PCTL^{*}$ and $\PCTL$
equivalences, respectively. Section \ref{sec:weak}  extends them to weak
(branching) bisimulations, and Section \ref{sec:simulation} extends the framework to simulations. We discuss the extension to countable states in Section \ref{sec:countable} and the coarsest congruent bisimulations  and simulations in Section \ref{sec:congruent}. In Section \ref{sec:related} we discuss
related work, and Section \ref{sec:conclusion} concludes the paper.

\section{Preliminaries}\label{sec:pre}
\paragraph{Distributions.}
For a \emph{countable} set $S$, a distribution is a function $\mu:S\to
[0,1]$ satisfying $\ABS{\mu}:=\sum_{s\in S}\mu(s)= 1$. We denote by
$\mathit{Dist}(S)$ the set of distributions over $S$. We shall use
$s,r,t,\ldots$ and $\mu,\nu\ldots$ to range over $S$ and
$\mathit{Dist}(S)$, respectively. Given a set of distributions
$\{\mu_i\}_{1\leq i\leq n}$, and a set of positive weights
$\{w_i\}_{1\leq i\leq n}$ such that $\sum_{1\leq i\leq n}w_i=1$, the
\emph{convex combination} $\mu=\sum_{1\leq i\leq n}w_i\cdot\mu_i$ is
a distribution such that $\mu(s)=\sum_{1\leq i\leq
  n}w_i\cdot\mu_i(s)$ for each $s\in S$. The support of $\mu$ is
defined by $\mathit{supp}(\mu):=\{s\in S \mid \mu(s)>0\}$. For an
equivalence relation $\MC{R}$, we write $\mu~\MC{R}~\nu$ if it holds
that $\mu(C)=\nu(C)$ for all equivalence classes $C\in S/\MC{R}$. A
distribution $\mu$ is called \emph{Dirac} if $|\mathit{supp}(\mu)|=1$,
and we let $\DIRAC{s}$ denote the Dirac distribution with
$\DIRAC{s}(s)=1$.

\paragraph{Downward Closure.} We define the downward closure of
a set of states.
For a relation $\MC{R}$ over $S$ and $C\subseteq S$, define
$\DOWNWARD{\MC{R}}{(C)}$ as the least set satisfying:
i) $C\subseteq\DOWNWARD{\MC{R}}{(C)}$, ii) $(s,s')\in\MC{R}$ and
$s'\in \DOWNWARD{\MC{R}}{(C)}$ implies $s\in\DOWNWARD{\MC{R}}{(C)}$.
We say $C$ is $\MC{R}$ \emph{downward closed} iff $C=\DOWNWARD{\MC{R}}{(C)}$.
We use $\DOWNWARD{\MC{R}}{(s)}$ as the shorthand of $\DOWNWARD{\MC{R}}{(\{s\})}$, 
and $\DOWNWARD{\MC{R}}{}=\{C\mid C\subseteq S\land C=\DOWNWARD{\MC{R}}{(C)}\}$ to 
denote the set of all $\MC{R}$ downward closed sets. If $\MC{R}$ is an equivalence
relation, then $C$ is called $\MC{R}$ closed if $C=\DOWNWARD{\MC{R}}(C)$.

\section{Probabilistic automata, \texorpdfstring{$\PCTL^{*}$}{PCTL*} and bisimulations}\label{sec:pa}
\subsection{Probabilistic automata}
We recall the notion of probabilistic automata introduced by Segala~\cite{Segala-thesis}.  We omit the set of actions, since they do not
appear in the logic $\PCTL$ we shall consider later. This is actually not
a restriction, since the bisimulation we shall introduce later
can be extended to $\PA$s with
actions directly.
\begin{defi}\label{def:automata}
  A \emph{probabilistic automaton} is a tuple
  $\mathcal{P}=(S,\rightarrow,s_0,\AP,L)$ where $S$ is a finite set of
  states, $\rightarrow~\subseteq S\times\MI{Dist}(S)$ is a
  transition relation such that for each $s\in S$, 
  there exists $(s,\mu)\in\rightarrow$ for some $\mu$, 
  $s_0\in S$ is the initial state,
  $\AP$ is a set of atomic propositions, and $L:S\rightarrow
  2^{\mathit{AP}}$ is a labeling function.
\end{defi}
We only consider image-finite $\PA$s, i.e.
$\{\mu\mid (s,\mu)\in\rightarrow\}$ is finite for each $s\in S$. A transition
$(s,\mu)\in\rightarrow$ is often denoted by $\TRANA{s}{}{\mu}$. Moreover, we
write $\TRANA{\mu}{}{\mu'}$ iff for each $s\in\mathit{supp}(\mu)$
there exists $\TRANA{s}{}{\mu_s}$ such that
$\mu'(r)=\sum\limits_{s\in\mathit{supp}(\mu)}\mu(s)\cdot\mu_s(r)$.

A \emph{path} is a finite or infinite sequence
$\omega=s_0s_1s_2\ldots$ of states. 
We use $\mathit{lstate}(\omega)$ and $l(\omega)$ to denote the last
state of $\omega$ and the length of $\omega$ respectively if $\omega$
is finite. The set $\mathit{Path}$ is the set containing all paths, and
$\mathit{Path}(s_0)$ contains those starting from $s_0$. Similarly,
$\mathit{Path}^{*}$ is the set of finite paths, and
$\mathit{Path}^{*}(s_0)$ only contains finite paths starting from $s_0$. Also we use
$\omega[i]$ to denote the $(i+1)$-th state for $i\geq 0$, $\omega|^i$
to denote the prefix of $\omega$ ending at $\omega[i]$, and
$\omega|_i$ to denote the suffix of $\omega$ starting from
$\omega[i]$.

We introduce the definition of \emph{schedulers} to resolve non-determinism.
A scheduler is a
function $\sigma:\mathit{Path}^{*}\mapsto
\mathit{Dist}(\rightarrow)$ such that
$\sigma(\omega)(s,\mu)>0$ implies
$s=\mathit{lstate}(\omega)$. A scheduler $\sigma$ is
\emph{deterministic} if it returns only Dirac distributions, that
is, the next step is chosen deterministically.

The \emph{cone} of a finite path $\omega$, denoted by $C_{\omega}$,
is the set of paths having $\omega$ as their prefix, i.e.,
$C_{\omega}=\{\omega'\mid\omega\leq\omega'\}$ where $\omega\leq\omega'$ iff $\omega$ is a prefix of $\omega'$.
Fixing a starting state $s_0$ and a scheduler $\sigma$, the measure
$\MEASURE_{\sigma,s_0}$ of a cone $C_{\omega}$, where
$\omega=s_0s_1\ldots s_k$, is defined inductively as follows: $\MEASURE_{\sigma,s_0}(C_{\omega})$ equals $1$ if $k=0$, and for $k>0$,
$$
\MEASURE_{\sigma,s_0}(C_{\omega})=\MEASURE_{\sigma,s_0}(C_{\omega|^{k-1}})\cdot\left(\sum\limits_{(s_{k-1},\mu')\in\rightarrow}\sigma(\omega|^{k-1})(s_{k-1},\mu')\cdot\mu'(s_k)\right)
$$

Let $\mathcal{B}$ be the smallest algebra that
contains all the cones and is closed under complement and countable
unions. By standard measure theory~\cite{halmos1974measure,rudin2006real} this algebra is a
$\sigma$\emph{-algebra} and all its elements are
measurable sets of paths. Given a scheduler $\sigma$,
 $\MEASURE_{\sigma,s_0}$ can be extended to a unique
measure on $\mathcal{B}$.

Given a relation $\MC{R}$ over $S$, $(\DOWNWARD{\MC{R}}{})^i$ is the \emph{Cartesian} product of
$\DOWNWARD{\MC{R}}{}$ with itself $i$ times.
Each element of $(\DOWNWARD{\MC{R}}{})^i$  is a \emph{downward closed path} of length $i$.
Let $(\DOWNWARD{\MC{R}}{})^+=\mathop{\cup}_{i\geq
  1}(\DOWNWARD{\MC{R}}{})^i$, and define $l(\Omega)=n$ for $\Omega\in(\DOWNWARD{\MC{R}}{})^n$.
For $\Omega=C_0C_1\ldots C_n\in(\DOWNWARD{\MC{R}}{})^+$, the
$\MC{R}$ \emph{downward closed cone} $C_{\Omega}$ is defined as
$C_{\Omega}=\{C_{\omega}\mid\omega\in\Omega\}$,
where $\omega\in\Omega$ iff $\omega[i]\in C_i$ for $0\leq i\leq n$.

For distributions $\mu_1$ and $\mu_2$, we define $\mu_1\times \mu_2$ by $(\mu_1\times\mu_2)((s_1,s_2))=\mu_1(s_1)\times\mu_2(s_2)$.
Following~\cite{baier2008principles} we also define the interleaving of $\PA$s:
\begin{defi}\label{def:interleave}
Let $\MC{P}_i=(S_i,\rightarrow_i,s_i,\AP_i,L_i)$ be two $\PA$s with $i=1,2$. The \emph{interleaved parallel composition} $\MC{P}_1\interleave\MC{P}_2$ is defined by:
\[\MC{P}_1\interleave\MC{P}_2 = (S_1\times S_2,\rightarrow, (s_1,s_2),\AP_1\times\AP_2,L)\]
where $L((r_1,r_2))=L_1(r_1)\times L_2(r_2)$ and $\TRAN{(r_1,r_2)}{\mu}$ iff either $\TRAN{r_1}{\mu_1}$ and $\mu=\mu_1\times\DIRAC{r_2}$, or $\TRAN{r_2}{\mu_2}$ and $\mu=\DIRAC{r_1}\times\mu_2$.
\end{defi}

\subsection{\texorpdfstring{$\PCTL^{*}$}{PCTL*} and its sub logics}
We introduce the syntax of $\PCTL$~\cite{hansson1994logic} and $\PCTL^{*}$~\cite{Aziz1995UWT} which are probabilistic extensions of {\sf CTL} and {\sf CTL}$^{*}$ respectively.

The $\PCTL^{*}$  formulas over the set $\AP$ of atomic propositions are
formed according to the following grammar:
\begin{align*}
\phi &::= a \mid \phi_1\wedge\phi_2\mid\neg \phi\mid\MC{P}_{\bowtie q}(\psi)\\
\psi &::=\phi\mid\psi_1\land\psi_2\mid\neg\psi\mid\X\psi\mid\psi_1\U\psi_2
\end{align*}
where $a\in\AP$, $\bowtie\ \in\{<,>,\leq,\geq\}$, and $q\in[0,1]$. We refer to $\phi$ and $\psi$ as ($\PCTL^{*}$) state and
path formulas, respectively.

The satisfaction relation $s\models \phi$ for state formulas is
 defined in a standard manner for boolean formulas. For the probabilistic operator, it is defined by
$$s\models\MC{P}_{\bowtie q}(\psi)\text{ iff }\forall
\sigma.\MEASURE_{\sigma,s}(\{\omega\in\mathit{Path}(s)\mid\omega\models\psi\})\bowtie q.$$
The satisfaction relation $\omega\models\psi$ for path
formulas is the same as for LTL formulas, that is,
\begin{align*}
&\omega\models\X\psi &&\text{ iff }  \omega|_1\models\psi\\
&\omega\models\psi_1\U\psi_2 && \text{ iff }   \exists j\geq
0.\omega|_j\models\psi_2\land\forall 0\leq k<j.\omega|_k\models\psi_1
\end{align*}

\paragraph{Sub logics.}
The depth of path formula $\psi$ of $\PCTL^{*}$ free of $\U$ operator, denoted by $\DEPTH(\psi)$, is defined by the maximum number of embedded $\X$ operators appearing in $\psi$, that is,
\begin{iteMize}{$\bullet$}
\item $\DEPTH(\phi)=0$,
\item $\DEPTH(\psi_1\land\psi_2)=\max\{\DEPTH(\psi_1),\DEPTH(\psi_2)\}$,
\item $\DEPTH(\neg\psi)=\DEPTH(\psi)$ and
\item $\DEPTH(\X\psi)=1+\DEPTH(\psi)$.
\end{iteMize}
Then, we let
$\PCTL^{*-}$ be the sub logic of $\PCTL^{*}$ without the until ($\psi_1\U\psi_2$) operator. Moreover, $\PCTL^{*-}_i$ is a sub logic of $\PCTL^{*-}$ where
 for each $\psi$ we have $\DEPTH(\psi)\leq i$.

The sub logic $\PCTL$ is obtained by restricting the path formulas to:
\[\psi::=\X\phi\mid\phi_1\U\phi_2 \mid\phi_1\U^{\le n}\phi_2\]
Note the bounded until operator does not appear in $\PCTL^{*}$ as it
can be encoded by nested next operators.  $\PCTL^{-}$ is defined in a
similar way as $\PCTL^{*-}$.  Moreover we let $\PCTL^{-}_i$ be the
sub logic of $\PCTL^{-}$ where only bounded until operator
$\phi_1\U^{\leq j}\phi_2$ with $j\leq i$ is allowed. For all the logics
we have mentioned, we summarize their
differences in syntax of path formulas in
Table~\ref{tab:logics}.

\begin{table}
\caption{Summary of $\PCTLS$ and its sublogics}\label{tab:logics}
\centering
\begin{tabular}{|l|c|c|}
\hline
$\qquad$Logic$\qquad$ & $\qquad\psi\qquad$ & $\qquad$Note$\qquad$\\[3pt]
\hline
$\PCTL^*$ &  $\phi\mid\psi_1\land\psi_2\mid\neg\psi\mid\X\psi\mid\psi_1\U\psi_2$  &\\[3pt]
\hline
$\PCTL^{*-}$ & $\phi\mid\psi_1\land\psi_2\mid\neg\psi\mid\X\psi$ &\\[3pt]
\hline
$\PCTL^{*-}_i$ &$\phi\mid\psi_1\land\psi_2\mid\neg\psi\mid\X\psi$& $\DEPTH(\psi)\leq i$\\[3pt]
\hline
$\PCTL$ & $\X\phi\mid\phi_1\U\phi_2 \mid\phi_1\U^{\le n}\phi_2$ &\\[3pt]
\hline
$\PCTL^{-}$ & $\X\phi\mid\phi_1\U^{\le n}\phi_2$ & \\[3pt]
\hline
$\PCTL^{-}_i$ & $\X\phi\mid\phi_1\U^{\le j}\phi_2$ & $j\leq i$\\[3pt]
\hline
\end{tabular}
\end{table}

\paragraph{Logical equivalence.}
For a logic $\mathcal{L}$,
we say that $s$ and $r$ are $\mathcal{L}$-equivalent, denoted by $s \sim_{\mathcal{L}} r$,
if they satisfy the same set of formulas of $\mathcal{L}$,
that is $s\models\phi$ iff $r\models\phi$ for all state formulas $\phi$ in $\mathcal{L}$.
The logic $\mathcal{L}$ can be $\PCTL^{*}$ or one of its sub logics.

\subsection{Strong probabilistic bisimulation}\label{sec:strong probabilistic bisimulation}
In this section we introduce the definition of strong probabilistic bisimulation~\cite{SegalaL95}.
Let $\{\TRAN{s}{\mu_i}\}_{i\in I}$ be a collection of
transitions of $\MC{P}$, and let $\{w_i\}_{i\in I}$ be a
collection of probabilities with $\sum_{i\in I}w_i=1$. Then
$(s,\sum_{i\in I}w_i\cdot\mu_i)$ is called a \emph{combined
transition} and is denoted by $\TRANP{s}{\mu}$ where
$\mu=\sum_{i\in I}w_i\cdot\mu_i$.

\begin{defi}\label{def:strong probabilistic bisimulation}
An equivalence relation $\MC{R}\subseteq S\times S$ is a strong
probabilistic bisimulation iff $s~\MC{R}~r$ implies that $L(s)=L(r)$ and for each $\TRAN{s}{\mu}$, there exists a combined transition
$\TRANP{r}{\mu'}$ such that $\mu~\MC{R}~\mu'$.

We write $s~\BSP~r$ whenever there is a strong
probabilistic bisimulation $\MC{R}$ such that $s~\MC{R}~r$.
\end{defi}

It was shown in~\cite{SegalaL95} that $\BSP$ is preserved by $\interleave$, that is, $s~\BSP~r$ implies $s\interleave t~\BSP~r\interleave t$ for any $t$. Also strong probabilistic bisimulation is sound for $\PCTL$ which means that if $s\ \BSP\ r$ then for any state formula $\phi$ of $\PCTL$, $s\models \phi$
iff $r\models \phi$. But the other way around is not true, i.e.
strong probabilistic bisimulation is not complete for $\PCTL$, as illustrated by
the following example.
\begin{exa}\label{ex:counterexample}
  Consider again the two $\PA$s in Fig.~\ref{fig:counterexample} and
  assume that all the states have different labels except $s$ and
  $r$. In addition, $s_1$, $s_2$, and $s_3$ only
  have one transition to themselves with probability 1.  The only
  difference between $s$ and $r$ is that $r$ has an extra step. We can see that the maximal/minimal
  probabilities from $s$ and $r$ to any set $C\subseteq \{s_1,s_2,s_3\}$
  are the same, therefore it holds that
  $s~\EPCTLS~r$. But by Definition~\ref{def:strong probabilistic
    bisimulation} we have $s~\nBSP~r$, because the middle transition
  of $r$ cannot be simulated by $s$ even if we use combined transitions
  i.e. there is no $a,b\in[0,1]$ such that $0.3\cdot a+0.5\cdot
  b=0.4$, $0.3\cdot a + 0.4\cdot b=0.3$, and $0.4\cdot a + 0.1\cdot b
  = 0.3$. Therefore we conclude that strong probabilistic bisimulation
  is not complete for $\PCTL^{*}$ as well as for $\PCTL$.
\end{exa}

It should be noted that $\PCTL^{*}$ distinguishes more states in a $\PA$ than $\PCTL$. Refer to the following example.
\begin{exa}\label{ex:pclt and pclt star}
  Suppose $s$ and $r$ are given by Fig.~\ref{fig:counterexample} where
  each of $s_1$, $s_2$, and $s_3$ is extended with a transition such
  that $\TRAN{s_1}{\mu_1}$ with $\mu_1(s_1)=0.6$ and $\mu_1(s_4)=0.4$,
  $\TRAN{s_2}{\mu_2}$ with $\mu_2(s_4)=1$, and $\TRAN{s_3}{\mu_3}$
  with $\mu_3(s_3)=0.5$ and $\mu_3(s_4)=0.5$. Here we assume that every state satisfies
  different atomic propositions except that $L(s)=L(r)$. Then it is not hard to
  see $s~\EPCTL~r$ while $s~\nEPCTLS~r$. Consider
the $\PCTL^{*}$  formula  $$\phi=\MC{P}_{\leq0.38}(\X(L(s_1)\lor
  L(s_3))\land\X\X(L(s_1)\lor L(s_3))),$$ it holds $s\models\phi$ but $r\not\models\phi$. Note that $\phi$ is not a
  well-formed $\PCTL$ formula. Indeed, states $s$ and $r$ are $\PCTL$-equivalent.
\end{exa}

We remark that $\CTL$ and $\CTL^{*}$ equivalences coincide on transition systems, but quantitative properties have more distinguishing power at this point.

We have the following theorem:
\begin{thm}\label{thm:PCLT and Probabilistic bisimulation}\hfill
\begin{enumerate}[\em(1)]
\item $\EPCTL$, $\EPCTLS$, $\EPCTLM$, $\iEPCTLM{i}$, $\EPCTLSM$, $\iEPCTLSM{i}$, and $\BSP$ are equivalence relations for any $i\geq 1$.
\item $\BSP~\subset~\EPCTLS~\subset~\EPCTL$.
\item $\EPCTLSM~\subset~\EPCTLM$.
\item $\iEPCTLSM{1}~=~\iEPCTLM{1}$.
\item $\iEPCTLSM{i}~\subset~\iEPCTLM{i}$ for any $i>1$.
\item $\EPCTL~\subset~\EPCTLM~\subset~\iEPCTLM{i+1}~\subset~\iEPCTLM{i}$ for all $i\geq 0$.
\item $\EPCTLS~\subset~\EPCTLSM~\subset~\iEPCTLSM{i+1}~\subset~\iEPCTLSM{i}$ for all $i\geq 0$.
\end{enumerate}
\end{thm}
\begin{proof}
We take $\EPCTL$ as an example and all the others can be proved in a similar way. The reflexivity is trivial. If $s~\EPCTL~r$, then we also have $r~\EPCTL~s$ since $s$ and $r$ satisfy the same set of formulas, hence we prove the symmetry of $\EPCTL$. Now we prove the transitivity, that is, for any $s,r,t$ if we have $s~\EPCTL~r$ and $r~\EPCTL~t$, then $s~\EPCTL~t$. It is also easy, since $s$ and $r$ satisfy the same set of formulas, and $r$ and $t$ satisfy the same set of formulas by $s~\EPCTL~r$ and $r~\EPCTL~t$, as a result $s\models\phi$ implies $t\models\phi$ and vice versa for any $\phi$, so $s~\EPCTL~t$. We conclude that $\EPCTL$ is an equivalence relation.

The proof of $\BSP~\subset~\EPCTL$ can be found in~\cite{SegalaL95} while $\BSP~\subset~\EPCTLS$ can be proved in a similar way.
The proof of $\EPCTLS~\subset~\EPCTL$ is trivial since $\PCTL$ is a subset of $\PCTL^{*}$. Example~\ref{ex:pclt and pclt star} shows that there exists
states which are $\PCTL$ equivalent, but not $\PCTLS$ equivalent, therefore
the inclusion is strict.

The proofs of Clause 3 and 5 are obvious since $\EPCTLM$ is a subset of $\EPCTLSM$ while $\iEPCTLM{i}$ is a subset of $\iEPCTLSM{i}$.

We now prove that $\iEPCTLSM{1}~=~\iEPCTLM{1}$. It is sufficient to prove that $\PCTL^{-}_1$ and $\PCTL^{*-}_1$ have the same expressiveness.
The proof of $\iEPCTLSM{1}~\subseteq~\iEPCTLM{1}$ is easy since $\PCTL^{-}_1$ is a subset of $\PCTL^{*-}_1$. We now show how formulas of $\PCTL^{*-}_1$ can be encoded by formulas of $\PCTL^{-}_1$. It is not hard to see that the syntax of path formulas of $\PCTL^{*-}_1$ can be rewritten as:
\[\psi::=\phi\mid\X\phi\mid\neg\psi\mid\psi_1\land\psi_2\]
where we replace $\X\psi$ with $\X\phi$ since $\PCTL^{*-}_1$ only allows path formulas
whose depths are less or equal than 1. Since $\neg\X\phi=\X\neg\phi$, therefore we only need to consider
the following cases: $\MC{P}_{\bowtie q}(\phi)$, $\MC{P}_{\bowtie q}(\X\phi_1\land\X\phi_2)$, $\MC{P}_{\bowtie q}(\X\phi_1\land\phi_2)$,
and $\MC{P}_{\bowtie q}(\neg\psi)$. We distinguish several cases:
\begin{enumerate}[(1)]
\item $0<q\leq 1$ and $\bowtie = \geq$:
\begin{enumerate}
\item $s\models\MC{P}_{\geq q}(\phi)$ iff $s\models\phi$;
\item $s\models\MC{P}_{\geq q}(\X\phi_1\land\X\phi_2)$ iff $s\models\MC{P}_{\geq q}(\X(\phi_1\land\phi_2))$;
\item $s\models\MC{P}_{\geq q}(\X\phi_1\land\phi_2)$ iff $s\models\phi_2\land\MC{P}_{\geq q}(\X\phi_1)$;
\item $s\models\MC{P}_{\geq q}(\neg\psi)$ iff $s\models\MC{P}_{\leq (1-q)}(\psi)$.
\end{enumerate}
\item $q=0$ and $\bowtie = \geq$:\\
This case is trivial, since $s\models\MC{P}_{\geq 0}(\psi)$ iff $s\models\top$ for
any $\psi$ where $\top=a\lor\neg a$ for some $a$.
\item $0\leq q < 1$ and $\bowtie = \leq$:
\begin{enumerate}
\item $s\models\MC{P}_{\leq q}(\phi)$ iff $s\not\models\phi$;
\item $s\models\MC{P}_{\leq q}(\X\phi_1\land\X\phi_2)$ iff $s\models\MC{P}_{\leq q}(\X(\phi_1\land\phi_2))$;
\item $s\models\MC{P}_{\leq q}(\X\phi_1\land\phi_2)$ iff $s\models\neg\phi_2\lor\MC{P}_{\leq q}(\X\phi_1)$;
\item $s\models\MC{P}_{\leq q}(\neg\psi)$ iff $s\models\MC{P}_{\geq (1-q)}(\psi)$.
\end{enumerate}
\item $q=1$ and $\bowtie = \leq$: \\
Similar as Clause (2), $s\models\MC{P}_{\leq 1}(\psi)$ iff $s\models\top$ for any
$\psi$.
\item The cases when $\bowtie=>$ or $<$ are similar and omitted here.
\end{enumerate}

The proofs of Clauses 6 and 7 are straightforward.
\end{proof}

\section{A novel strong bisimulation}\label{sec:strong}
This section presents our main contribution of the paper: We introduce a novel notion of  strong bisimulation and strong branching bisimulation. We shall show that they agree with  $\PCTL$ and $\PCTL^{*}$ equivalences, respectively.
As a preparation step we introduce the strong $1$-depth bisimulation.

\subsection{Strong \texorpdfstring{$1$}{1}-depth bisimulation}\label{sec:1 depth bisimulation}
\begin{defi}\label{def:1 depth bisimualtion}
A relation $\MC{R}\subseteq S\times S$ is a
strong $1$-depth bisimulation if $s~\MC{R}~r$ implies that $L(s)=L(r)$ and for any $\MC{R}$ downward closed set $C$
\begin{enumerate}[(1)]
\item for each $\TRAN{s}{\mu}$, there exists $\TRAN{r}{\mu'}$ such that $\mu'(C)\geq\mu(C)$,
\item for each $\TRAN{r}{\mu}$, there exists $\TRAN{s}{\mu'}$ such that $\mu'(C)\geq\mu(C)$.\medskip
\end{enumerate}

\noindent We write $s~\iBS{1}~r$ whenever there is a strong $1$-depth bisimulation $\MC{R}$ such that $s~\MC{R}~r$.
\end{defi}

The -- though very simple -- definition  requires only one step matching
of the distributions out of $s$ and $r$. The essential
difference to the standard definition is: The quantification of the downward closed set
comes before the quantification over transition. This is indeed the key of the new definition of bisimulation. The following theorem shows that $\iBS{1}$ agrees with $\iEPCTLM{1}$ and $\iEPCTLSM{1}$ which is also an equivalence relation:

\begin{lem}\label{thm:1 equivalence}
$\iEPCTLM{1}~=~\iBS{1}~=~\iEPCTLSM{1}$.
\end{lem}
\begin{proof}
  According to Clause (4) of Theorem~\ref{thm:PCLT and Probabilistic
    bisimulation}, it is enough to prove that
  $\iEPCTLM{1}~=~\iBS{1}$. Refer to the proof of
  Theorem~\ref{thm:relation of equivalence branching} for the details.
\end{proof}

Note that in Definition~\ref{def:1 depth bisimualtion} we consider all the $\MC{R}$ downward closed sets since it is not enough to only consider $\MC{R}$ downward closed sets in $\{\DOWNWARD{\MC{R}}{(s)}\mid s\in S\}$. Refer to the following example:
\begin{exa}\label{cex:all closed sets}
Suppose there are four absorbing states $s_1,s_2,s_3,$ and $s_4$ with different 
atomic propositions, and we have two processes $s$ and $r$ such that $L(s)=L(r)$, 
and $\TRAN{s}{\mu_1}$, $\TRAN{s}{\mu_2}$, $\TRAN{r}{\nu_1}$, $\TRAN{r}{\nu_2}$ 
where $\mu_1(s_1)=0.5$, $\mu_1(s_2)=0.5$, $\mu_2(s_3)=0.5$, $\mu_2(s_4)=0.5$, 
$\nu_1(s_1)=0.5$, $\nu_1(s_3)=0.5$, $\nu_2(s_2)=0.5$, $\nu_2(s_4)=0.5$.
Let $\MC{R}=\{(s,r)\}\cup\MI{ID}$ where $\MI{ID}$ denote the identical relation.
If we only consider $\MC{R}$ downward closed sets in $\{\DOWNWARD{\MC{R}}{(s)}\mid s\in S\}$ where $S=\{s,r,s_1,s_2,s_3,s_4\}$, then we will conclude that $s~\iBS{1}~r$, but $r\models\phi$ while $s\not\models\phi$ where $\phi=\MC{P}_{\geq0.5}(\X(L(s_1)\lor L(s_2)))$.
\end{exa}

It turns out that $\iBS{1}$ is preserved by $\interleave$, implying that $\iEPCTLM{1}$ and $\iEPCTLSM{1}$ are preserved by $\interleave$ as well.
\begin{lem}\label{lem:1-composition}
$s~\iBS{1}~r$ implies that $s\interleave t~\iBS{1}~r\interleave t$ for any $t$.
\end{lem}

\proof
Let $\MC{R}=\{(\PAR{s}{t},\PAR{r}{t})\mid s~\iBS{1}~r\}$, it is enough to show that $\MC{R}$
is a strong 1-depth bisimulation. Suppose $s~\iBS{1}~r$, and there exists $\TRAN{\PAR{s}{t}}{\mu}$ such
that $\mu(C)>0$ for a $\MC{R}$ downward closed set $C$. We need to show that there exists
$\TRAN{\PAR{r}{t}}{\mu'}$ such that $\mu'(C)\geq\mu(C)$.
By the definition of $\interleave$ operator, if $\TRAN{\PAR{s}{t}}{\mu}$,
then either $\TRAN{s}{\mu_s}$ with $\mu=\mu_s\interleave\DIRAC{t}$,
or $\TRAN{t}{\mu_t}$ with $\mu=\DIRAC{s}\interleave\mu_t$.
We only consider the case when $\mu=\mu_s\interleave\DIRAC{t}$, since the other one is similar. According to the definition of $\MC{R}$, for each $\MC{R}$ 
downward closed set $C$, there exists a $\iBS{1}$ downward closed set
$C'$ such that $\mu(C)=\mu(\{\PAR{s'}{t}\mid s'\in C'\})=\mu_s(C')$.
We have known that $s~\iBS{1}~r$, so for each $\TRAN{s}{\mu_s}$ and $C'$,
there exists $\TRAN{r}{\mu_r}$ such that $\mu_r(C')\geq\mu_s(C')$.
Therefore for each $C$ and $\TRAN{s\interleave t}{\mu}$,
there exists $\TRAN{r\interleave t}{\mu'}$ such that $$\mu'(C)=\mu'(\{\PAR{s'}{t}\mid s'\in C'\})= \mu_r(C') \geq \mu_s(C') = \mu(\{\PAR{s'}{t}\mid s'\in C'\})=\mu(C).\eqno{\qEd}$$\medskip

\noindent A few remarks are in order.
\begin{enumerate}[(1)]
\item Note in Definition 5 of~\cite{SongZG2011} we require that $\MC{R}$ is a preorder, while the $\MC{R}$ in Definition~\ref{def:1 depth bisimualtion} can be any relation, we could also have required $\MC{R}$ being an equivalence relation: but all of them will induce the same largest bisimulation equivalence.

  Bisimulation relations are defined for arbitrary relations for
  classical transition
  systems~\cite{milner1989communication}. However, in the literature
  of bisimulation relations for probabilistic systems, bisimulation
  relations are defined mostly only for equivalence relations, see for
  example~\cite{LarsenS89,larsen1991bisimulation,Segala-thesis,baier2003comparative,SegalaL95,prakash-book}. For
  probabilistic systems, defining bisimulations for equivalence
  relations has the advantage of being very easy to understand. On
  the other side, a general definition for all relations is important
  as well, as it particularly sheds light to the connections to the
  classical transition systems. To the best of our knowledge, such
  general bisimulation definitions are first defined, independently,
  in~\cite{DesharnaisLT08,lpar08-yuxin-van-glabbeek,lijun-thesis}. Later
  in~\cite{Parma2007LCB,HermannsPSWZ11,SackZ12,Hennessy12}, this general definition
  has been exploited to study logical characterizations, and characterizing formulas 
  for probabilistic systems.
\item We note that for Kripke structures ($\PA$s with only Dirac
  distributions) $\iBS{1}$ agrees with the usual strong bisimulation
  by Milner~\cite{milner1989communication} if we consider
  state-labelled instead of transition-labelled systems.
\end{enumerate}
\subsection{Strong branching bisimulation}\label{sec:i depth bisimulation branching}
Now we extend the relation $\iBS{1}$ to strong $i$-step bisimulation. Then, the intersection of all of these relations gives us the new notion of strong branching bisimulation, which is shown to be the same as  $\EPCTL$. Recall that Theorem~\ref{thm:PCLT and Probabilistic bisimulation} states that  $\EPCTL$ is strictly coarser than $\EPCTLS$, which we shall consider in the next section.

Following the approach in~\cite{van1996branching}, we define $\MEASURE_{\sigma,s}(C,C',n,\omega)$ which denotes the probability from $s$ to states in $C'$ via states in $C$ possibly in at most $n$ steps under scheduler $\sigma$, where $\omega$ is used to keep track of the path and only deterministic schedulers are considered in the following. Formally, $\MEASURE_{\sigma,s}(C,C',n,\omega)$ equals $1$ if $s\in C'$, and else if $n>0\land(s\in C\setminus C')$, then
\begin{equation}\label{eq:definition of transition branching}
\MEASURE_{\sigma,s}(C,C',n,\omega)=
\sum\limits_{r\in\mathit{supp}(\mu')}\mu'(r)\cdot\MEASURE_{\sigma,r}(C,C',n-1,\omega r).
\end{equation}
where $\sigma(\omega)(s,\mu')=1$, otherwise $\MEASURE_{\sigma,s}(C,C',n,\omega)$ equals 0, provided $n>0$.

Strong $i$-depth branching bisimulation is a straightforward extension of strong $1$-depth bisimulation, where instead of considering only one immediate step, we consider up to $i$ steps. We let $\iBSB{1} ~=~\iBS{1}$ in the following.
\begin{defi}\label{def:index strong bisimulation branching}
A relation $\MC{R}\subseteq S\times S$ is a
strong $i$-depth branching bisimulation with $i>1$ if $s~\MC{R}~r$ implies $s~\iBSB{i-1}~r$ and for any $\MC{R}$ downward closed sets $C,C'$,
\begin{enumerate}[(1)]
\item for each scheduler $\sigma$, there exists a scheduler $\sigma'$ such that $$\MEASURE_{\sigma',r}(C,C',i,r)\geq\MEASURE_{\sigma,s}(C,C',i,s),$$
\item for each scheduler $\sigma$, there exists a scheduler $\sigma'$ such that $$\MEASURE_{\sigma',s}(C,C',i,s)\geq\MEASURE_{\sigma,r}(C,C',i,r).$$
\end{enumerate}
We write $s~\iBSB{i}~r$ whenever there is a strong $i$-depth branching bisimulation $\MC{R}$ such that $s~\MC{R}~r$. The strong branching bisimulation $\iBSB{}$ is defined as $\iBSB{}~=~\cap_{i\ge 1}\iBSB{i}$.
\end{defi}

Obviously, if $\MC{R}$ is symmetric, the second condition can be dropped.
The following lemma shows that $\iBSB{i}$ is an equivalence relation, and moreover,  $\iBSB{i}$ decreases until a fixed point is reached.
\begin{lem}\label{lem:i equivalence relation branching}\hfill
\begin{enumerate}[\em(1)]
\item $\iBSB{}$ and $\iBSB{i}$ are  equivalence relations for any $i\ge 1$.
\item $\iBSB{j}~\subseteq~\iBSB{i}$ provided that $1\leq i \leq j$.
\item There exists $i\geq 1$ such that $\iBSB{j}~=~\iBSB{k}$ for any $j,k\geq i$.
\end{enumerate}
\end{lem}
\begin{proof}
  We only show the proof of transitivity of $\iBSB{i}$. Suppose that
  $s~\iBSB{i}~t$ and $t~\iBSB{i}~r$, we need to prove that
  $s~\iBSB{i}~r$. By Definition~\ref{def:index strong bisimulation
    branching}, we know there exists strong $i$-depth branching
  bisimulations $\MC{R}_1$ and $\MC{R}_2$ such that $s~\MC{R}_1~t$ and
  $t~\MC{R}_2~r$. Assume $\MC{R}_1$ and $\MC{R}_2$ are reflexive, such
  relations always exist since each state is strong $i$-depth bisimilar to
  itself.
  Let $\MC{R}=\MC{R}_1\circ\MC{R}_2=\{(s_1,s_3)\mid\exists
  s_2.(s_1~\MC{R}_1~s_2\land s_2~\MC{R}_2~r)\},$ it is enough to show
  that $\MC{R}$ is a strong $i$-depth branching bisimulation. Note
  $\MC{R}_1\cup\MC{R}_2~\subseteq~\MC{R}$, since for each
  $s_1~\MC{R}_1~s_2$ we also have $s_2~\MC{R}_2~s_2$ due to
  reflexivity, thus $s_1~\MC{R}~s_2$, implying $\MC{R}_1\subseteq\MC{R}$.
  Similarly we can show that
  $\MC{R}_2~\subseteq~\MC{R}$. Therefore for any $\MC{R}$ downward
  closed sets $C$ and $C'$, they are also $\MC{R}_1$ and $\MC{R}_2$
  downward closed. Therefore for each $\sigma$, 
  there exists $\sigma'$ such that
  $\MEASURE_{\sigma',t}(C,C',i,t)\geq\MEASURE_{\sigma,s}(C,C',i,s)$. Since
  $t~\iBSB{i}~r$, thus there exists $\sigma''$ such that
  $\MEASURE_{\sigma'',r}(C,C',i,r)\geq\MEASURE_{\sigma',t}(C,C',i,t)\geq\MEASURE_{\sigma,s}(C,C',i,s)$. This
  completes the proof of transitivity.

The proof of Clause (2) is straightforward from Definition~\ref{def:index strong bisimulation branching}, since $s~\iBSB{j}~r$ implies $s~\iBSB{j-1}~r$ when $j>1$.

From Definition~\ref{def:index strong bisimulation branching}, we can see that $\iBSB{i}$ is getting more discriminating as $i$ increases.
Moreover, in a $\PA$ only with finitely many states the maximum number of equivalence classes is equal to the number of states, as result we can guarantee that $\iBSB{n}~=~\iBSB{}$ where $n$ is the total number of states.
\end{proof}

Given a relation $\MC{R}$, two paths $\omega_1=s_0s_1\ldots$, $\omega_2=r_0r_1\ldots$ are in $\MC{R}$, written as $\omega_1~\MC{R}~\omega_2$, iff $\omega_1[j]~\MC{R}~\omega_2[j]$ for all $j\geq 0$.
We then define the $\MC{R}$ closed paths:
The set $\Omega$ of paths is $\MC{R}$ \emph{closed} if for any
$\omega_1\in\Omega$ and $\omega_2$ such that $\omega_1~\MC{R}~\omega_2$,
it holds that $\omega_2\in\Omega$. Let
$\mathcal{B}_{\MC{R}}=\{\Omega\subseteq\mathcal{B}\mid
\Omega\text{ is $\MC{R}$ closed}\}$. By standard measure theory
$\mathcal{B}_{\MC{R}}$ is measurable.

In the following, we will use $\mathit{Sat}(\phi)=\{s\in S\mid s\models \phi\}$ to denote the set of states which satisfy $\phi$.
Similarly, $\mathit{Sat}(\psi)=\{\omega\in\mathit{Path}\mid\omega\models\psi\}$ is the set of paths satisfying $\psi$.
Below we show that $\iBSB{i}$  characterizes $\PCTL^{-}_i$. Moreover, we show that $\iBSB{}$ agrees with $\PCTL$ equivalence.
\begin{thm}\label{thm:relation of equivalence branching}
$\iEPCTLM{i}~=~\iBSB{i}$ for any $i\geq 1$, moreover  $\EPCTL\ =\ \iBSB{}$.
\end{thm}

\proof\hfill
\begin{enumerate}[(1)]
\item $\iEPCTLM{i}~\subseteq~\iBSB{i}$:\\
Let $\MC{R}=\{(s,r)\mid s~\iEPCTLM{i}~r\}$ and
  $s~\MC{R}~r$. Thus, $\MC{R}$ is symmetric. We show that for any
  $\MC{R}$ closed sets $C,C'$ and scheduler $\sigma$, there exists a
  scheduler $\sigma'$ such that
  $\MEASURE_{\sigma',r}(C,C',i,r)\geq\MEASURE_{\sigma,s}(C,C',i,s)$.
  Suppose there are $n$ different equivalence classes of $\MC{R}$ 
  in a finite $\PA$. Let
  $\phi_{C_i,C_j}$ be a state formula such that
  $\mathit{Sat}(\phi_{C_i,C_j})\supseteq C_i$ and
  $\mathit{Sat}(\phi_{C_i,C_j})\cap C_j=\emptyset$, here $1\leq i\neq
  j\leq n$ and $C_i,C_j\in S/\MC{R}$ are two different equivalence
  classes. Formula like $\phi_{C_i,C_j}$ always exists, otherwise
  there will not exist a formula which is fulfilled by states in
  $C_i$, but not fulfilled by states in $C_j$, that is, states in
  $C_i$ and $C_j$ satisfy the same set of formulas, this is against
  the assumption that $C_i$ and $C_j$ are two different equivalence
  classes. Let $\phi_{C_i}=\mathop{\land}\limits_{1\leq j\neq i\leq
    n}\phi_{C_i,C_j}$, it is not hard to see that
  $\mathit{Sat}(\phi_{C_i})=C_i$. For a $\MC{R}$ closed set $C$ which is
  a set of equivalence classes, let
  $\phi_C=\mathop{\vee}\limits_{C'\in S/\MC{R}\land
    C'\subseteq C}\phi_{C'},$ then it holds $\mathit{Sat}(\phi_C)=C$. Now
  suppose $\MEASURE_{\sigma,s}(C,C',i,s)=q$, then we
  know $s\models\neg\MC{P}_{<q}(\psi)$ where $\psi=\phi_{C}\U^{\leq
    i}\phi_{C'}$. By assumption $r\models\neg\MC{P}_{<q}(\psi)$, so there
  exists a scheduler $\sigma'$ such that
  $\MEASURE_{\sigma',r}(C,C',i,r)\geq q$, that is,
  $\MEASURE_{\sigma',r}(C,C',i,r)\geq\MEASURE_{\sigma,s}(C,C',i,s)$.
\item $\iBSB{i}~\subseteq~\iEPCTLM{i}$:\\
The proof is done by structural induction on the syntax of state formula $\phi$ and path formula $\psi$ of $\PCTL^{-}_i$, that is, we need to prove the following two results simultaneously.
\begin{enumerate}
\item $s~\iBSB{i}~r$ implies that $s\models\phi$ iff $r\models\phi$ for any state formula $\phi$ of $\PCTL_i$.
\item $\omega_1~\iBSB{i}~\omega_2$ implies that $\omega_1\models\psi$ iff $\omega_2\models\psi$ for any path formula $\psi$ of $\PCTL_i$.
\end{enumerate}
We only consider $\phi=\MC{P}_{\leq q}(\psi)$ where
$\psi=\phi_1\U^{\leq i}\phi_2$, since other cases are
similar. According to the semantics $s\models\phi$ iff
$\forall\sigma.\MEASURE_{\sigma,s}(\{\omega\mid\omega\models\psi\})\leq
q$. Since $\psi=\phi_1\U^{\leq i}\phi_2$, we only need to consider
prefix of length $i$ for each path. 
By induction hypothesis $\{\omega\mid\omega\models\psi\}$ is $\iBSB{i}$ closed.
Since $\psi=\phi_1\U^{\leq i}\phi_2$, there exists
$\Omega$ such 
that $l(\Omega)\leq i$ and $C_\Omega=\{\omega\mid\omega\models\psi\}$. 
According to the semantics of $\U^{\leq i}$, 
there exists two $\iBSB{i}$ closed sets $C,C'$
such that $\Omega=\mathop{\cup}\limits_{0\leq k<i}C^kC'$. 
We prove by contradiction, and assume $s\models\phi$ and $r\not\models\phi$. Then
for any $\sigma$, we have $\MEASURE_{\sigma,s}(C_\Omega)\leq q$. Since
$r\not\models\phi$, there exists $\sigma'$ such that
$\MEASURE_{\sigma',r}(C_\Omega)>q$, thus there does not exist $\sigma$
such that
$\MEASURE_{\sigma,s}(C,C',i,s)\geq\MEASURE_{\sigma',r}(C,C',i,r)$,
which contradicts the assumption $s~\iBSB{i}~r$. Therefore
$r\models\phi$, and $s~\iEPCTLM{i}~r$.
\item $\EPCTL~\subseteq~\iBSB{}$:\\
Since $\EPCTL~\subseteq~\iEPCTLM{i}$ for each $i\ge 0$, thus $\EPCTL~\subseteq~\cap_{i\ge 0}\iEPCTLM{i}$. According to above proof, 
$\EPCTL~\subseteq~\cap_{i\ge 0}\iEPCTLM{i}~=~\cap_{i\ge 0}\iBSB{i}~=~\iBSB{}$.
\item $\iBSB{}~\subseteq~\EPCTL$:\\
For any scheduler $\sigma$, we have 
$$\MEASURE_{\sigma,s}(\{\omega\mid\omega\models\phi_1\U^{\le i}\phi_2\}) = \MEASURE_{\sigma,s}(\MI{Sat}(\phi_1),\MI{Sat}(\phi_2),i,s).$$ 
Moreover $
\MEASURE_{\sigma,s}(\{\omega\mid\omega\models\phi_1\U\phi_2\})
=\lim_{i\rightarrow\infty}\MEASURE_{\sigma,s}(\MI{Sat}(\phi_1),\MI{Sat}(\phi_2),i,s),
$
the limit exists since $\{\MEASURE_{\sigma,s}(\MI{Sat}(\phi_1),\MI{Sat}(\phi_2),i,s)\mid i\ge 0\}$ is a monotonic and bounded i.e. convergent sequence. Therefore $s~\iEPCTLM{i}~r$ (or $s~\iBSB{i}~r$) for any $i\ge 0$ 
implies that $s~\EPCTL~r$.\qed 
\end{enumerate}

Intuitively, since $\iBSB{i}$ becomes smaller as $i$ increases, for any $\PA$, $\iBSB{i}$ will eventually converge to $\PCTL$ equivalence.

Recall $\iBSB{1}$ is compositional by Lemma~\ref{lem:1-composition}, which unfortunately is not the case for $\iBSB{i}$ with $i>1$. This is illustrated by the following example:

\begin{counter}\label{cex:i compositional branching}
$s~\iBSB{i}~r$ does not necessarily imply $s\interleave t~\iBSB{i}~r\interleave t$ for any $t$ generally if $i>1$.

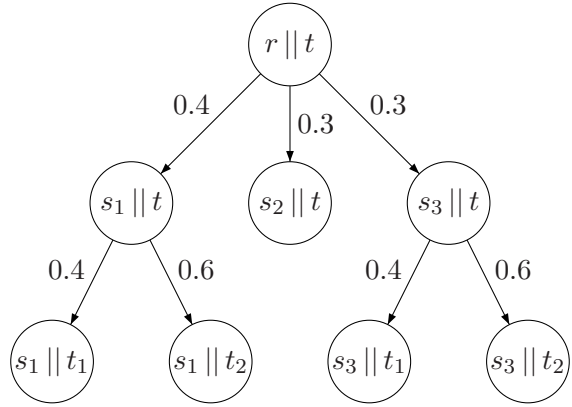
\begin{figure}[!t]
\begin{center}
    \begin{picture}(100, 50)(-50,-50)
    \gasset{Nw=11,Nh=11,Nmr=5.5,curvedepth=0}
    \unitlength=3pt
    \node(A)(0,0){$\PAR{r}{t}$}
    \node(B)(-20,-20){$\PAR{s_1}{t}$}
    \node(C)(0,-20){$\PAR{s_2}{t}$}
    \node(D)(20,-20){$\PAR{s_3}{t}$}
    \node(E)(-30,-40){$\PAR{s_1}{t_1}$}
    \node(F)(-10,-40){$\PAR{s_1}{t_2}$}
    \node(G)(10,-40){$\PAR{s_3}{t_1}$}
    \node(H)(30,-40){$\PAR{s_3}{t_2}$}
    \drawedge[ELside=r](A,B){0.4}
    \drawedge(A,C){0.3}
    \drawedge(A,D){0.3}
    \drawedge[ELside=r](B,E){0.4}
    \drawedge(B,F){0.6}
    \drawedge[ELside=r](D,G){0.4}
    \drawedge(D,H){0.6}
    \end{picture}
  \end{center}
  \caption{\label{fig:non-composition} $\iBSB{i}$ is not compositional when $i>1$}
\end{figure}

We have shown in Example~\ref{ex:counterexample} that $s~\EPCTL~r$. If
we compose $s$ and $r$ with $t$ where $t$ only has a transition to
$\mu$ such that $\mu(t_1)=0.4$ and $\mu(t_2)=0.6$, then it turns out
that $s\interleave t~\nEPCTL~r\interleave t$. Since there exists
$\phi=\MC{P}_{\leq 0.34}(\psi)$  with \[\psi=((L(s\interleave t)\lor
L(s_1\interleave t)\lor(L(s_3\interleave t)))\U^{\leq 2}
(L(s_1\interleave t_2)\lor L(s_3\interleave t_1)))\] such that
$s\interleave t\models\phi$ but $r\interleave t \not\models \phi$, as
there exists a scheduler $\sigma$ such that the probability of paths
satisfying $\psi$ in $\MEASURE_{\sigma,r}$ equals
0.36. Fig.~\ref{fig:non-composition} shows the execution of $\PAR{r}{t}$ guided
by $\sigma$, where we assume all the states in
Fig.~\ref{fig:non-composition} have different atomic propositions
except that $L(s\interleave t)=L(r\interleave t)$. It is similar for
$\EPCTLS$.

Note that $\phi$ is also a well-formed state formula of $\PCTL^{-}_{2}$, so $\iEPCTLM{i}$ as well as $\iBSB{i}$ are not compositional if $i\geq 2$.
\end{counter}

\subsection{Strong bisimulation}\label{sec:i depth bisimulation}
In this section we introduce a new notion of strong bisimulation and show that it
characterizes $\EPCTLS$. Given  a relation  $\MC{R}$, a $\MC{R}$ downward closed cone $C_{\Omega}$ and
a measure $\MEASURE$, the value of
$\MEASURE(C_{\Omega})$ can be computed by summing up the values of all
$\MEASURE(C_{\omega})$ with $\omega\in\Omega$. We let
$\tilde{\Omega}\subseteq(\DOWNWARD{\MC{R}}{})^+$ be a set of
$\MC{R}$ \emph{downward closed paths}, then $C_{\tilde{\Omega}}$ is the
corresponding set of $\MC{R}$ downward closed cones, that is,
$C_{\tilde{\Omega}}=\mathop{\cup}_{\Omega\in\tilde{\Omega}}C_{\Omega}$. Define $l(\tilde{\Omega})=\mathit{Max}\{l(\Omega)\mid\Omega\in\tilde{\Omega}\}$
as the maximum length of $\Omega$ in $\tilde{\Omega}$. To
compute $\MEASURE(C_{\tilde{\Omega}})$, we cannot sum up the value of each
$\MEASURE(C_{\Omega})$ such that $\Omega\in\tilde{\Omega}$ as before, since we
may have a path $\omega$ such that $\omega\in\Omega_1$ and
$\omega\in\Omega_2$ where $\Omega_1,\Omega_2\in\tilde{\Omega}$, so we
have to remove these duplicate paths and make sure each path is
considered once and only once as follows,
where we abuse the notation and write $\omega\in\tilde{\Omega}$ iff $\exists\Omega\in\tilde{\Omega}.\omega\in\Omega$:
\begin{equation}\label{eq:set of closed cones}
\MEASURE(C_{\tilde{\Omega}}) = \mathop{\sum}\limits_{\omega\in\tilde{\Omega}\land\not\exists\omega'\in\tilde{\Omega}.\omega'\leq\omega}\MEASURE(C_{\omega})
\end{equation}
Note Equation~\ref{eq:set of closed cones}  can be extended to compute the probability of any set of cones in a given measure.

The definition of strong $i$-depth bisimulation is as follows:
\begin{defi}\label{def:index strong bisimulation}
A relation $\MC{R}\subseteq S\times S$ is a
strong $i$-depth bisimulation if $i>1$ and $s~\MC{R}~r$ implies that $s~\iBS{i-1}~r$ and for any $\tilde{\Omega}\subseteq (\DOWNWARD{\MC{R}}{})^+$ with $l(\tilde{\Omega})=i$
\begin{enumerate}[(1)]
\item for each scheduler $\sigma$, there exists $\sigma'$ such that $\MEASURE_{\sigma',r}(C_{\tilde{\Omega}})\geq\MEASURE_{\sigma,s}(C_{\tilde{\Omega}})$,
\item for each scheduler $\sigma$, there exists $\sigma'$ such that $\MEASURE_{\sigma',s}(C_{\tilde{\Omega}})\geq\MEASURE_{\sigma,r}(C_{\tilde{\Omega}})$.
\end{enumerate}

We write $s~\iBS{i}~r$ whenever there is a $i$-depth strong bisimulation $\MC{R}$ such that $s~\MC{R}~r$. The strong bisimulation $\iBS{}$ is defined as $\iBS{}\ =\ \cap_{i\ge 1} \iBS{i}$.
\end{defi}

Recall that $(\DOWNWARD{\MC{R}}{})^+$ contains all the downward
closed paths. Each downward closed path can be equivalently stated as
a sequence of downward closed sets, thus Definition~\ref{def:index
  strong bisimulation} subsumes Definition~\ref{def:index strong
  bisimulation branching} in the sense that the two downward closed
sets $C$ and $C'$ in Definition~\ref{def:index strong bisimulation
  branching} can be seen as a special downward closed path of form
$CC\ldots C'$. Similar to $\iBSB{i}$, the relation $\iBS{i}$ forms a
chain of equivalence relations, and $\iBS{i}$ will converge finally in
a $\PA$.

\begin{lem}\label{lem:i equivalence relation}\hfill
\begin{enumerate}[\em(1)]
\item $\iBS{i}$ is an equivalence relation for any $i> 1$.
\item $\iBS{j}~\subseteq~\iBS{i}$ provided that $1\leq i \leq j$.
\item There exists $i\geq 1$ such that $\iBS{j}~=~\iBS{k}$ for any $j,k\geq i$.
\end{enumerate}
\end{lem}
\begin{proof}
For the first clause we only prove the transitivity since the reflexivity and
  symmetry are easy. Suppose that $s~\iBS{i}~r$ and $r~\iBS{i}~t$, we
  need to show that $s~\iBS{i}~t$. According to
  Definition~\ref{def:index strong bisimulation}, we know there exists
  strong $i$-depth bisimulations $\MC{R}_1$ and $\MC{R}_2$ such that
  $s~\MC{R}_1~t$ and
  $t~\MC{R}_2~r$. Let $\MC{R}=\MC{R}_1\circ\MC{R}_2=\{(s_1,s_3)\mid\exists
  s_2.(s_1~\MC{R}_1~s_2\land s_2~\MC{R}_2~r)\},$ it is enough to show
  that $\MC{R}$ is a strong $i$-depth bisimulation. Similar as in the
  proof of Lemma~\ref{lem:i equivalence relation branching}, if
  $\tilde{\Omega}\subseteq (\DOWNWARD{\MC{R}}{})^+$, then it also
  holds that $\tilde{\Omega}\subseteq (\DOWNWARD{\MC{R}_1}{})^+$ and
  $\tilde{\Omega}\subseteq (\DOWNWARD{\MC{R}_2}{})^+$. Thus for each
  $\tilde{\Omega}\subseteq (\DOWNWARD{\MC{R}}{})^+$ with
  $l(\tilde{\Omega})=i$, and scheduler $\sigma$ of $s$, there exists
  $\sigma'$ of $r$ such that
  $\MEASURE_{\sigma',r}(\tilde{\Omega})\geq\MEASURE_{\sigma,s}(\tilde{\Omega})$. Since
  $r~\iBS{i}~t$, there exists scheduler $\sigma''$ of $t$ such that
    $$\MEASURE_{\sigma'',t}(\tilde{\Omega})\geq\MEASURE_{\sigma',r}(\tilde{\Omega})\geq\MEASURE_{\sigma,s}(\tilde{\Omega}).$$ The other direction is similar and omitted here, thus $s~\iBS{i}~t$.

The proof for the second clause  is straightforward from Definition~\ref{def:index strong bisimulation}. For the last one,
since there are only finitely many states, thus there are only
  finitely many equivalence classes, such $i$ always exists.
\end{proof}

Below we show that $\iBS{i}$ characterizes $\iEPCTLSM{i}$ for all $i\ge 1$, where $\infBS=\mathop{\cap}\limits_{n\geq 1}\iBS{n}$.
\begin{thm}\label{thm:relation of equivalence}
$\iEPCTLSM{i}~=~\iBS{i}$ for any $i\geq1$, moreover $\EPCTLS~=~\iBS{}$.
\end{thm}

\proof\hfill
\begin{enumerate}[(1)]
\item $\iEPCTLSM{i}~\subseteq~\iBS{i}$:\\
Let $\MC{R}=\{(s,r)\mid s~\iEPCTLSM{i}~r\}$ and $s~\MC{R}~r$, obviously $\MC{R}$ is symmetric. We show that for any $\tilde{\Omega}\subseteq (\DOWNWARD{\MC{R}}{})^+$ with $l(\tilde{\Omega})=i$ and scheduler $\sigma$, there exists a scheduler $\sigma'$ such that $\MEASURE_{\sigma',r}(C_{\tilde{\Omega}})\ge\MEASURE_{\sigma,s}(C_{\tilde{\Omega}})$.
Following the way in the proof of Theorem~\ref{thm:relation of equivalence branching}, we can construct a formula $\phi_C$ such that $\mathit{Sat}(\phi_C)=C$ where $C$ is a $\MC{R}$ closed set. Suppose $\Omega=C_0C_1\ldots C_j$ with $j\leq i$, then
\[\psi_{\Omega}=\phi_{C_0}\land\X(\phi_{C_1}\land\ldots\land\X(\phi_{C_{j-1}}\land\X\phi_{C_j})\ldots)\]
can be used to characterize $\Omega$, that is, $\mathit{Sat}(\psi_{\Omega})=C_{\Omega}$. Let $\psi=\mathop{\lor}\limits_{\Omega\in\tilde{\Omega}}\psi_{\Omega}$, then $\mathit{Sat}(\psi)=C_{\tilde{\Omega}}$. As a result $s\models\neg \MC{P}_{<q}(\psi)$ where $q=\MEASURE_{\sigma,s}(C_{\tilde{\Omega}})$. By assumption $r\models\neg \MC{P}_{<q}(\psi)$, so there exists a scheduler $\sigma'$ such that $\MEASURE_{\sigma',r}(C_{\tilde{\Omega}})\geq q$, that is, $\MEASURE_{\sigma',r}(C_{\tilde{\Omega}})\geq\MEASURE_{\sigma,s}(C_{\tilde{\Omega}})$.
\item $\iBS{i}~\subseteq~\iEPCTLSM{i}$:\\
The proof is by structural induction on the syntax of state formula $\phi$ and path formula $\psi$ of $\PCTL^{*-}_i$, that is, we need to prove the following two results simultaneously.
\begin{enumerate}
\item $s~\iBS{i}~r$ implies that $s\models\phi$ iff $r\models\phi$ for any state formula $\phi$ of $\PCTL^{*-}_i$.
\item $\omega_1~\iBS{i}~\omega_2$ implies that $\omega_1\models\psi$ iff $\omega_2\models\psi$ for any path formula $\psi$ of $\PCTL^{*-}_i$.
\end{enumerate}
We only consider $\phi=\MC{P}_{\leq q}(\psi)$ such that $\DEPTH(\psi)\leq i$. 
By induction hypothesis $\{\omega\mid\omega\models\psi\}$ is $\iBS{i}$ closed. Since
$\DEPTH(\psi) \leq i$, there exists $\tilde{\Omega}$ such that
$l(\tilde{\Omega})\leq i$ and 
$C_{\tilde{\Omega}}=\{\omega\mid\omega\models\psi\}$.
We prove by contradiction, and assume that $s\models\phi$ and $r\not\models\phi$. According to the semantics $s\models\phi$ iff $\forall\sigma.\MEASURE_{\sigma,s}(C_{\tilde{\Omega}})\leq q$. If $r\not\models\phi$, then there exists $\sigma'$ such that $\MEASURE_{\sigma',r}(C_{\tilde{\Omega}})>q$, consequently for such $\sigma'$ of $r$ there does not exist $\sigma$ of $s$ such that $\MEASURE_{\sigma,s}(C_{\tilde{\Omega}})\geq\MEASURE_{\sigma',r}(C_{\tilde{\Omega}})$ which contradicts the assumption that $s~\iBS{i}~r$, therefore $r\models\phi$ and $s~\iEPCTLSM{i}~r$.
\item $\EPCTLS~=~\iBS{}$:\\
The arguments are similar as in Theorem~\ref{thm:relation of equivalence branching}.\qed
\end{enumerate}


\noindent For the same reason as strong $i$-depth branching bisimulation,  $\iBS{i}$ is not preserved by $\interleave$ when $i>1$.
\begin{counter}\label{cex:i compositional}
$s~\iBS{i}~r$ does not necessarily imply $s\interleave t~\iBS{i}~r\interleave t$ for any $t$ generally if $i>1$.
This can be shown by using the same arguments as in Counterexample~\ref{cex:i compositional branching}.
\end{counter}

\subsection{Taxonomy for strong bisimulations}
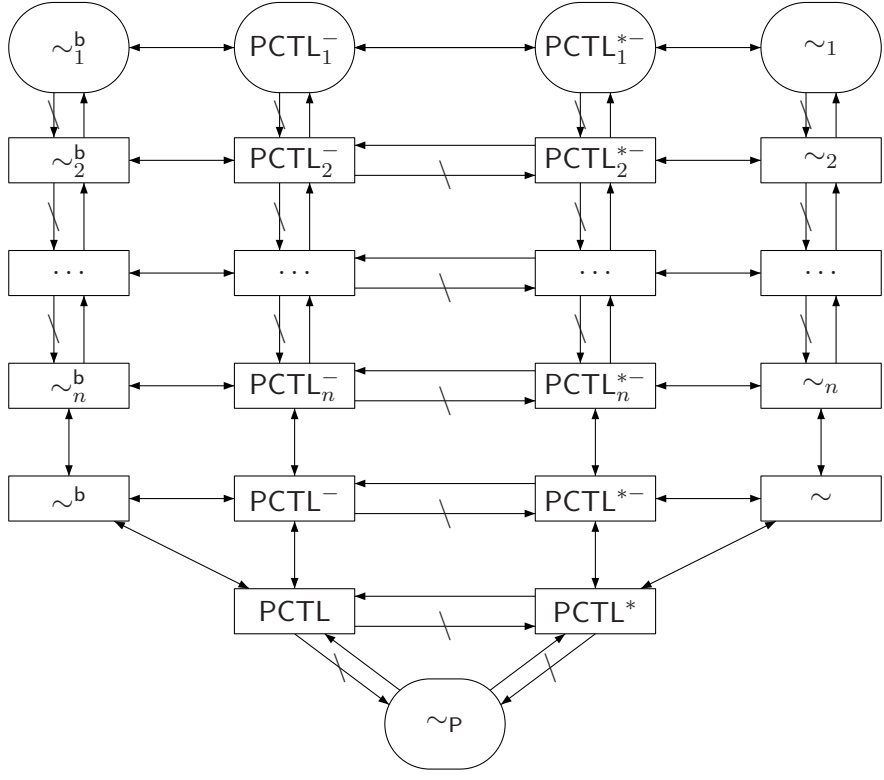
\begin{figure}[!t]
  \begin{center}
  \begin{picture}(60, 100)(-30,-10)
  \gasset{Nw=16,Nh=6,Nmr=0}
  \node(P)(-20,15){$\PCTL$}
  \node(PM)(-20,30){$\PCTL^{-}$}
  \node(PNM)(-20,45){$\PCTL^{-}_n$}
  \node(DT)(-20,60){$\ldots$}
  \node(PTM)(-20,75){$\PCTL^{-}_2$}
  \node(PS)(20,15){$\PCTL^{*}$}
  \node(PSM)(20,30){$\PCTL^{*-}$}
  \node(PNSM)(20,45){$\PCTL^{*-}_n$}
  \node(DTH)(20,60){$\ldots$}
  \node(PTSM)(20,75){$\PCTL^{*-}_2$}
  \node(BS)(50,30){$\infBS$}
  \node(IBSN)(50,45){$\iBS{n}$}
  \node(DF)(50,60){$\ldots$}
  \node(IBST)(50,75){$\iBS{2}$}
  \node(BSB)(-50,30){$\infBSB$}
  \node(IBSBN)(-50,45){$\iBSB{n}$}
  \node(DO)(-50,60){$\ldots$}
  \node(IBSBT)(-50,75){$\iBSB{2}$}
  \gasset{Nw=16,Nh=12,Nmr=6}
  \node(BSP)(0,0){$\BSP$}
  \node(POM)(-20,90){$\PCTL_1^{-}$}
  \node(IBSBO)(-50,90){$\iBSB{1}$}
  \node(IBSO)(50,90){$\iBS{1}$}
  \node(POSM)(20,90){$\PCTL^{*-}_1$}
  \drawedge[AHnb=1,ATnb=1](IBSBO,POM){}
  \drawedge[AHnb=1,ATnb=1](POSM,POM){}
  \drawedge[AHnb=1,ATnb=1](IBSO,POSM){}
  \drawedge[AHnb=1,ATnb=1](IBSBT,PTM){}
  \drawedge[AHnb=1,ATnb=1](IBSBN,PNM){}
  \drawedge[AHnb=1,ATnb=1](BSB,PM){}
  \drawedge[AHnb=1,ATnb=1](DO,DT){}
  \drawedge[AHnb=1,ATnb=1](BSB,P){}
  \drawedge[AHnb=1,ATnb=1](IBST,PTSM){}
  \drawedge[AHnb=1,ATnb=1](IBSN,PNSM){}
  \drawedge[AHnb=1,ATnb=1](BS,PSM){}
  \drawedge[AHnb=1,ATnb=1](DTH,DF){}
  \drawedge[AHnb=1,ATnb=1](BS,PS){}
  \drawedge[ELside=r,exo=2,sxo=2](IBSBT,IBSBO){}
  \drawedge[ELside=r,exo=2,sxo=2](DO,IBSBT){}
  \drawedge[ELside=r,exo=2,sxo=2](IBSBN,DO){}
  \drawedge[AHnb=1,ATnb=1](BSB,IBSBN){}
  \drawedge[ELside=r,exo=2,sxo=2](PTM,POM){}
  \drawedge[ELside=r,exo=2,sxo=2](DT,PTM){}
  \drawedge[ELside=r,exo=2,sxo=2](PNM,DT){}
  \drawedge[AHnb=1,ATnb=1](PM,PNM){}
  \drawedge[AHnb=1,ATnb=1](P,PM){}
  \drawedge(BSP,PS){}
  \drawedge[ELside=r,exo=2,sxo=2](IBST,IBSO){}
  \drawedge[ELside=r,exo=2,sxo=2](DF,IBST){}
  \drawedge[ELside=r,exo=2,sxo=2](IBSN,DF){}
  \drawedge[AHnb=1,ATnb=1](BS,IBSN){}
  \drawedge[ELside=r,exo=2,sxo=2](PTSM,POSM){}
  \drawedge[ELside=r,exo=2,sxo=2](DTH,PTSM){}
  \drawedge[ELside=r,exo=2,sxo=2](PNSM,DTH){}
  \drawedge[AHnb=1,ATnb=1](PSM,PNSM){}
  \drawedge[AHnb=1,ATnb=1](PS,PSM){}
  \drawedge(BSP,P){}
  \drawedge(POSM,POM){}
  \drawedge[eyo=2,syo=2](PTSM,PTM){}
  \drawedge[eyo=2,syo=2](DTH,DT){}
  \drawedge[eyo=2,syo=2](PNSM,PNM){}
  \drawedge[eyo=2,syo=2](PSM,PM){}
  \drawedge[eyo=2,syo=2](PS,P){}
  \gasset{ELdistC=y,ELdist=0}
  \drawedge[ELside=r,exo=-4,sxo=-4](P,BSP){$\setminus$}
  \drawedge[ELside=r,exo=4,sxo=4](PS,BSP){$\setminus$}
  \drawedge[ELside=r,exo=-2,sxo=-2,ELpos=60](IBSBO,IBSBT){$\setminus$}
  \drawedge[ELside=r,exo=-2,sxo=-2,ELpos=50](IBSBT,DO){$\setminus$}
  \drawedge[ELside=r,exo=-2,sxo=-2,ELpos=50](DO,IBSBN){$\setminus$}
  \drawedge[ELside=r,exo=-2,sxo=-2,ELpos=60](IBSO,IBST){$\setminus$}
  \drawedge[ELside=r,exo=-2,sxo=-2,ELpos=50](IBST,DF){$\setminus$}
  \drawedge[ELside=r,exo=-2,sxo=-2,ELpos=50](DF,IBSN){$\setminus$}
    \drawedge[ELside=r,exo=-2,sxo=-2,ELpos=60](POM,PTM){$\setminus$}
  \drawedge[ELside=r,exo=-2,sxo=-2,ELpos=50](PTM,DT){$\setminus$}
  \drawedge[ELside=r,exo=-2,sxo=-2,ELpos=50](DT,PNM){$\setminus$}
    \drawedge[ELside=r,exo=-2,sxo=-2,ELpos=60](POSM,PTSM){$\setminus$}
  \drawedge[ELside=r,exo=-2,sxo=-2,ELpos=50](PTSM,DTH){$\setminus$}
  \drawedge[ELside=r,exo=-2,sxo=-2,ELpos=50](DTH,PNSM){$\setminus$}
  \drawedge[eyo=-2,syo=-2](PTM,PTSM){$\setminus$}
  \drawedge[eyo=-2,syo=-2](DT,DTH){$\setminus$}
  \drawedge[eyo=-2,syo=-2](PNM,PNSM){$\setminus$}
  \drawedge[eyo=-2,syo=-2](PM,PSM){$\setminus$}
  \drawedge[eyo=-2,syo=-2](P,PS){$\setminus$}
  \end{picture}
  \end{center}
  \caption{\label{fig:summary of relation} Relationship of different equivalences in strong scenario.}
\end{figure}

Fig.~\ref{fig:summary of relation} summaries the relationship among
all these bisimulations and logical equivalences, where $\rightarrow$
denotes $\subseteq$ and $\nrightarrow$ denotes $\nsubseteq$. We also
abbreviate $\EPCTL$ as $\PCTL$, and similarly for other logical
equivalences. The parameter $n$ means that for any finite $\PA$,
 there always exists a $n\ge 0$
such that $\iEPCTLM{n}~=~\iBSB{n}$ and $\iEPCTLSM{n}~=~\iBS{n}$.
Congruent relations with respect to the $\interleave$
operator are shown in circles, and non-congruent relations are shown
in boxes. Segala and Lynch have considered another strong bisimulation
in~\cite{SegalaL95}, which can be defined by replacing the
$\TRANP{r}{\mu'}$ with $\TRAN{r}{\mu'}$ in Definition~\ref{def:strong
  probabilistic bisimulation} and thus is strictly stronger than
$\BSP$.  It is also worth mentioning that all the bisimulations shown
in Fig.~\ref{fig:summary of relation} coincide with the strong
bisimulation defined in~\cite{BaierKHW05} in the {\sf DTMC}s setting
($\PA$s without non-deterministic choices). 

\section{Weak bisimulations}\label{sec:weak}
As in~\cite{BaierKHW05} we use $\PCTL_{\backslash\X}$ to denote the
subset of $\PCTL$ without the next operator $\X\phi$ and the bounded until
$\phi_1\U^{\leq n}\phi_2$. Similarly, $\PCTL^{*}_{\backslash\X}$ is
used to denote the subset of $\PCTL^{*}$ without the next operator
$\X\psi$. In this section we shall introduce  weak bisimulations and study
their relation to $\EPCTLWN$ and $\EPCTLSWN$, respectively. Before this we should
point out that $\EPCTLSWN$ implies $\EPCTLWN$ but the other direction
does not hold. Refer to the following example.

\begin{exa}\label{ex:counterexample weak}
  Suppose $s$ and $r$ are given by Fig.~\ref{fig:counterexample} where
  each of $s_1$ and $s_3$ is attached with one transition
  respectively, that is, $\TRAN{s_1}{\mu_1}$ such that
  $\mu_1(s_4)=0.4$ and $\mu_1(s_5)=0.6$, $\TRAN{s_3}{\mu_3}$ such that
  $\mu_3(s_4)=0.4$ and $\mu_3(s_5)=0.6$. In addition, $s_2$, $s_4$ and
  $s_5$ only have a transition with probability $1$ to themselves, and
  all these states are assumed to have different atomic
  propositions. Then $s~\EPCTLWN~r$ \footnote{This can be obtained by
    combining Theorem~\ref{thm:weak bisimulation and PCTL branching}
    and Example~\ref{ex:weak branching bisimulation} in
    Section~\ref{sec:weak bisimulation branching}.} but
  $s~\nEPCTLSWN~r$, since we have a path formula $$\psi=((L(s)\lor
  L(s_1))\U L(s_5))\lor((L(s)\lor L(s_3))\U L(s_4))$$ such that
  $s\models\MC{P}_{\leq0.34}(\psi)$ but $r\not\models\MC{P}_{\leq
    0.34}(\psi)$, since there exists a scheduler $\sigma$ where the
  probability of paths satisfying $\psi$ in
  $\MEASURE_{\sigma,r}$ is equal to
  $\MEASURE_{\sigma,r}(C_{ss_1s_5})+\MEASURE_{\sigma,r}(C_{ss_3s_4})=0.36$.
  Note $\psi$ is not a  $\PCTL_{\backslash\X}$ path formula.
\end{exa}

In this section we shall introduce a notion of branching bisimulation. 
Similar to the definition of bounded reachability $\MEASURE_{\sigma,s}(C,C',n,\omega)$, we first define the function $\MEASURE_{\sigma,s}(C,C',\omega)$ which denotes the probability
to go from $s$ to states in $C'$ possibly via states in $C$. Again $\omega$ is used to keep track of the path which has been visited. Formally,  $\MEASURE_{\sigma,s}(C,C',\omega)$ is equal to
$\MEASURE_{\sigma,s}(C,C',n,\omega)$ when $n\rightarrow\infty$, i.e.,
\begin{equation}\label{eq:definition of transition branching unbounded}
\MEASURE_{\sigma,s}(C,C',\omega)=\lim_{n\rightarrow\infty}\MEASURE_{\sigma,s}(C,C',n,\omega).
\end{equation}

\subsection{Weak probabilistic bisimulation}\label{sec:weak branching segala}
Before introducing our weak bisimulation, we give the classical
definition of weak probabilistic bisimulation proposed
in~\cite{SegalaL95}. Given an equivalence relation $\MC{R}$, $s$ can
evolve into $\mu$ by a \emph{weak branching transition}, written as
$\bTRAN{s}{\mu}$, iff there exists a scheduler $\sigma$ such that
$\mu(C)=\MEASURE_{\sigma,s}([s],C,s)$ for each $C\in S/\MC{R}$,
where $[s]$ denotes the equivalence class containing $s$. Intuitively, $\bTRAN{s}{\mu}$ means that $s$ can evolve into $\mu$
only via states in $[s]$. Accordingly, \emph{weak branching combined
  transition} $\bTRANP{s}{\mu}$ can be defined based on the weak branching
transition, i.e. $\bTRANP{s}{\mu}$ iff there exists a collection of
weak branching transitions $\{\bTRAN{s}{\mu_i}\}_{i\in I}$, and a
collection of probabilities $\{w_i\}_{i\in I}$ such that 
$\sum_{i\in I}w_i=1$ and $\mu=\sum_{i\in I}w_i\cdotp\mu_i$.

We give the definition of weak probabilistic bisimulation as follows:
\begin{defi}\label{def:branching bisimulation}
An equivalence relation $\MC{R}\subseteq S\times S$ is a weak probabilistic bisimulation iff $s~\MC{R}~r$ implies that $L(s)=L(r)$ and for each $\TRAN{s}{\mu}$, there exists $\bTRANP{r}{\mu'}$ such that $\mu~\MC{R}~\mu'$.

We write $s~\bBSP~r$ whenever there is a weak probabilistic bisimulation $\MC{R}$ such that $s~\MC{R}~r$.
\end{defi}

The following properties concerning weak probabilistic bisimulation are taken from~\cite{SegalaL95}:
\begin{lem}[\cite{SegalaL95}]\label{thm:branching and branching probabilistic}\hfill
\begin{enumerate}[\em(1)]
\item $\bBSP~\subset~\EPCTLSWN~\subset~\EPCTLWN$.
\item $\bBSP$ is preserved by $\interleave$.
\end{enumerate}
\end{lem}

\subsection{A novel branching bisimulation}\label{sec:weak bisimulation branching}
Below follows the definition of our branching bisimulation.
\begin{defi}\label{def:weak bisimulation branching}
A relation $\MC{R}\subseteq S\times S$ is a branching bisimulation if $s~\MC{R}~r$ implies that $L(s)=L(r)$ and for any $\MC{R}$ downward closed sets $C,C'$
\begin{enumerate}[(1)]
\item for each scheduler $\sigma$, there exists $\sigma'$ such that $\MEASURE_{\sigma',r}(C,C',r)\geq\MEASURE_{\sigma,s}(C,C',s)$,
\item for each scheduler $\sigma$, there exists $\sigma'$ such that $\MEASURE_{\sigma',s}(C,C',s)\geq\MEASURE_{\sigma,r}(C,C',r)$.
\end{enumerate}

\noindent We write $s~\WBSB~r$ whenever there is a branching bisimulation $\MC{R}$ such that $s~\MC{R}~r$.
\end{defi}

The following theorem shows that $\WBSB$ is an equivalence relation. Also different from the strong cases where we use a series of equivalence relations to either characterize or approximate $\EPCTL$ and $\EPCTLS$, in the weak scenario we show that $\WBSB$ itself is enough to characterize $\EPCTLWN$. Intuitively this is because in $\EPCTLWN$ only the unbounded until operator is allowed in path formulas which means we abstract from the number of steps to reach certain states.

\begin{exa}\label{ex:weak branching bisimulation}
  Refer to $s$ and $r$ described in Example~\ref{ex:counterexample
    weak}, we can
  show that $\MC{R}=\{(s,r)\}\cup\MI{ID}$ is a branching
  bisimulation. The only non-trivial case is to show that $(s,r)$ satisfies the conditions of
   the relation. According to Definition~\ref{def:weak bisimulation
    branching}, it is enough to check that for all the possible $C$
  and $C'$, the value of $\MEASURE_{\sigma_m,r}(C,C',r)$ will not be
  greater or smaller than $\MEASURE_{\sigma_l,s}(C,C',s)$ and
  $\MEASURE_{\sigma_r,s}(C,C',s)$ at the same time, where $\sigma_m$
  is the scheduler of $r$ always choosing the middle transition, while
  $\sigma_l$ and $\sigma_r$ are schedulers of $s$ always choosing the
  left transition and the right transition of $s$ respectively.
\end{exa}

\begin{thm}\label{thm:weak bisimulation and PCTL branching}\hfill
\begin{enumerate}[\em(1)]
\item $\WBSB$ is an equivalence relation.
\item $\WBSB~=~\EPCTLWN$.
\end{enumerate}
\end{thm}
\begin{proof}
The proof of the first clause is along the same line as the proof of
  Clause (1) of Lemma~\ref{lem:i equivalence relation branching}.
For the second clause, let $\MC{R}=\{(s,r)\mid s~\EPCTLWN~r\}$ and $s~\MC{R}~r$, where $\MC{R}$ is obviously symmetric. We show that for any $\MC{R}$ closed sets $C,C'$ and scheduler $\sigma$ of $s$, there exists a scheduler $\sigma'$ of $r$ such that $\MEASURE_{\sigma',r}(C,C',r)\geq\MEASURE_{\sigma,s}(C,C',s)$. Following the way in the proof of Theorem~\ref{thm:relation of equivalence branching}, we can construct a formula $\phi_C$ such that $\mathit{Sat}(\phi_C)=C$ where $C$ is a $\MC{R}$ closed set. Let $\psi=\phi_C\U\phi_{C'}$, then it is not hard to see that $s\models\neg \MC{P}_{<q}(\psi)$ where $q=\MEASURE_{\sigma,s}(C,C',s)$. By assumption $r\models\neg\MC{P}_{<q}(\psi)$, so there exists a scheduler $\sigma'$ such that $\MEASURE_{\sigma',r}(C,C',r)\geq q$, that is, $\MEASURE_{\sigma',r}(C,C',r)\geq\MEASURE_{\sigma,s}(C,C',s)$.

The proof of $\WBSB~\subseteq~\EPCTLWN$ is by structural induction on the syntax of state formula $\phi$ and path formula $\psi$ of $\PCTL_{\backslash\X}$, that is, we need to prove the following two results simultaneously.
\begin{enumerate}[(1)]
\item $s~\WBSB~r$ implies that $s\models\phi$ iff $r\models\phi$ for any state formula $\phi$.
\item $\omega_1~\WBSB~\omega_2$ implies that $\omega_1\models\psi$ iff $\omega_2\models\psi$ for any path formula $\psi$.
\end{enumerate}
We only consider $\phi=\MC{P}_{\leq q}(\psi)$ with
$\psi=\phi_1\U\phi_2$ since the other cases are similar.  By induction
hypothesis $\mathit{Sat}(\phi_1)$ and $\mathit{Sat}(\phi_2)$ are
$\WBSB$ closed, moreover
$\MEASURE_{\sigma,s}(\{\omega\mid\omega\models\psi\})=\MEASURE_{\sigma,s}(\mathit{Sat}(\phi_1),\mathit{Sat}(\phi_2),s)$
by Equation~(\ref{eq:definition of transition branching unbounded})
for any $\sigma$. We prove by contradiction, and assume that
$s\models\phi$ and $r\not\models\phi$. According to the semantics,
$s\models\phi$ iff
$\forall\sigma.\MEASURE_{\sigma,s}(\mathit{Sat}(\phi_1),\mathit{Sat}(\phi_2),s)\leq
q$. If $r\not\models\phi$, then there exists $\sigma'$ of $r$ such
that it holds
$\MEASURE_{\sigma',r}(\mathit{Sat}(\phi_1),\mathit{Sat}(\phi_2),r)>q$,
therefore for such $\sigma'$, there does not exist $\sigma$ of $s$
such that
$\MEASURE_{\sigma,s}(\mathit{Sat}(\phi_1),\mathit{Sat}(\phi_2),s)\geq
\MEASURE_{\sigma',r}(\mathit{Sat}(\phi_1),\mathit{Sat}(\phi_2),r)$,
which contradicts the assumption $s~\WBSB~r$. As a result, it must
hold that $r\models\phi$, and $s~\EPCTLWN~r$.
\end{proof}

As in the strong scenario, $\WBSB$ suffers from the same problem as $\iBSB{i}$ and $\iBS{i}$ with $i>1$, that is, it is not preserved by $\interleave$.
\begin{counter}\label{cex:weak bisimulation composition branching}
$s~\WBSB~r$ does not necessarily imply $s\interleave t~\WBSB~r\interleave t$ for any $t$.
This can be shown in a similar way as Counterexample~\ref{cex:i compositional branching},
since the result will still hold even if we replace the bounded until formula with an
unbounded until formula in Counterexample~\ref{cex:i compositional branching}.
\end{counter}

\subsection{Weak bisimulations}\label{sec:weak bisimulation}
In order to define weak bisimulation, we consider stuttering paths. Let $\Omega$ be a finite $\MC{R}$ downward closed path, then
\begin{equation}
C_{\Omega_{\mathit{st}}}=\begin{cases}C_{\Omega} & l(\Omega)=1\\ \mathop{\bigcup}\limits_{0\leq i<n.k_i\geq 0}C_{(\Omega[0])^{k_0}\ldots(\Omega[n-2])^{k_{n-2}}\Omega[n-1]} & l(\Omega)=n\geq 2\end{cases}
\end{equation}
is the set of $\MC{R}$ downward closed paths which contain all stuttering paths,
where $\Omega[i]$ denotes the $(i+1)$-th element in $\Omega$ such that $0\leq i<l(\Omega)$.
Accordingly, $C_{\tilde{\Omega}_{\mathit{st}}}=\mathop{\cup}\limits_{\Omega\in\tilde{\Omega}}C_{\Omega_{\mathit{st}}}$
contains all the stuttering paths of each $\Omega\in\tilde{\Omega}$.

Now we are ready to give the definition of weak bisimulation as follows:
\begin{defi}\label{def:weak bisimulation}
A relation $\MC{R}\subseteq S\times S$ is a
weak bisimulation if $s~\MC{R}~r$ implies that $L(s)=L(r)$ and for any $\tilde{\Omega}\subseteq (\DOWNWARD{\MC{R}}{})^+$
\begin{enumerate}[(1)]
\item for each scheduler $\sigma$, there exists $\sigma'$ such that  $\MEASURE_{\sigma',r}(C_{\tilde{\Omega}_{\mathit{st}}})\geq\MEASURE_{\sigma,s}(C_{\tilde{\Omega}_{\mathit{st}}})$,
\item for each scheduler $\sigma$, there exists $\sigma'$ such that $\MEASURE_{\sigma',s}(C_{\tilde{\Omega}_{\mathit{st}}})\geq\MEASURE_{\sigma,r}(C_{\tilde{\Omega}_{\mathit{st}}})$.
\end{enumerate}

\noindent We write $s~\WBS~r$ whenever there is a weak bisimulation $\MC{R}$ such that $s~\MC{R}~r$.
\end{defi}

The following theorem shows that $\WBS$ is an equivalence relation. For the same reason as in Theorem~\ref{thm:weak bisimulation and PCTL branching}, $\WBS$ is enough to characterize $\EPCTLSWN$ which gives us the following theorem.
\begin{thm}\label{thm:weak bisimulation and PCTL}\hfill
\begin{enumerate}[\em(1)]
\item $\WBS$ is an equivalence relation.
\item $\WBS~=~\EPCTLSWN$.
\end{enumerate}
\end{thm}
\begin{proof}
The proof of the first clause is along the
  same line as the proof of Clause (1) of Lemma~\ref{lem:i equivalence
    relation branching}.
For the second clause, let $\MC{R}=\{(s,r)\mid s~\EPCTLSWN~r\}$ and $s~\MC{R}~r$, thus $\MC{R}$ is a symmetric relation.
  It suffices to show that for any $\tilde{\Omega}\subseteq (\DOWNWARD{\MC{R}}{})^+$ and
  scheduler $\sigma$, there exists a scheduler $\sigma'$ such that
  $\MEASURE_{\sigma',r}(C_{\tilde{\Omega}_{\mathit{st}}})\geq\MEASURE_{\sigma,s}(C_{\tilde{\Omega}_{\mathit{st}}})$.
  Following the way in the
  proof of Theorem~\ref{thm:relation of equivalence branching}, we can
  construct a formula $\phi_C$ such that $\mathit{Sat}(\phi_C)=C$
  where $C$ is a $\MC{R}$ closed set. Let
  $\psi_{\Omega}=\phi_{C_0}\U(\phi_{C_1}\U\ldots\phi_{C_n})$ where
  $\Omega=C_{C_0\ldots C_n}$, then
  $\psi_{\tilde{\Omega}}=\mathop{\lor}\limits_{\Omega\in\tilde{\Omega}}\psi_{\Omega}$. It
  is easy to see that $s\models\neg\MC{P}_{<q}(\psi)$ where
  $q=\MEASURE_{\sigma,s}(C_{\tilde{\Omega}_{\mathit{st}}})$ and
  $\psi=\psi_{\tilde{\Omega}}$. By assumption $r\models\neg
  \MC{P}_{<q}(\psi)$, so there exists a scheduler $\sigma'$ such that
  $\MEASURE_{\sigma',r}(C_{\tilde{\Omega}_{\mathit{st}}})\geq q$, that
  is,
  $\MEASURE_{\sigma',r}(C_{\tilde{\Omega}_{\mathit{st}}})\geq\MEASURE_{\sigma,s}(C_{\tilde{\Omega}_{\mathit{st}}})$.

    The proof of $\WBS~\subseteq~\EPCTLSWN$ is by structural induction on the syntax of state formula $\phi$ and path formula $\psi$ of $\PCTL^{*}_{\backslash\X}$, that is, we need to prove the following two results simultaneously.
\begin{enumerate}[(1)]
\item $s~\WBS~r$ implies that $s\models\phi$ iff $r\models\phi$ for any state formula $\phi$.
\item $\omega_1~\WBS~\omega_2$ implies that $\omega_1\models\psi$ iff $\omega_2\models\psi$ for any path formula $\psi$.
\end{enumerate}

To make the proof clearer, we rewrite the syntax of $\PCTL^{*}_{\backslash\X}$ as follows which is equivalent to the original definition.
\[\psi::=\phi\mid\psi_1\lor\psi_2\mid\neg\psi\mid\psi_1\U\psi_2\]

We only consider $\phi'=\MC{P}_{\leq q}(\psi)$ here. It suffices to prove
that for each $\psi$, there exists
$\tilde{\Omega}\subseteq (\DOWNWARD{\WBS}{})^+$ such that
$C_{\tilde{\Omega}}=\mathit{Sat}(\psi)$. The
proof is by structural induction on $\psi$ as follows:
\begin{enumerate}[(1)]
\item $\psi=\phi$. By induction hypothesis $\mathit{Sat}(\phi)$ is $\WBS$ closed. Let $\tilde{\Omega}=\{\mathit{Sat}(\phi)\}$, then $C_{\tilde{\Omega}}=\mathit{Sat}(\psi)$.
\item $\psi=\psi_1\lor\psi_2$. By induction hypothesis there exist $\tilde{\Omega}'$ and $\tilde{\Omega}''$ such that $\mathit{Sat}(\psi_1)=C_{\tilde{\Omega}'_{\mathit{st}}}$ and $\mathit{Sat}(\psi_2)=C_{\tilde{\Omega}''_{\mathit{st}}}$. It is not hard to see that $\tilde{\Omega}=\tilde{\Omega}'\cup\tilde{\Omega}''$ will be enough.
\item $\psi=\psi_1\U\psi_2$. By induction hypothesis there exist $\tilde{\Omega}'$ and $\tilde{\Omega}''$ such that $\mathit{Sat}(\psi_1)=C_{\tilde{\Omega}'_{\mathit{st}}}$ and $\mathit{Sat}(\psi_2)=C_{\tilde{\Omega}''_{\mathit{st}}}$. Let $\tilde{\Omega}=\tilde{\Omega}''\cup\{\Omega'\Omega''\mid\Omega'\in\tilde{\Omega}'\land\Omega''\in\tilde{\Omega}''\}$, then $C_{\tilde{\Omega}}=\mathit{Sat}(\psi)$.
\item $\psi=\neg\psi'$. $s\models\MC{P}_{\geq q}(\psi)$ iff
  $s\models\MC{P}_{\le 1-q}(\psi')$, so $\psi$ can be reduced to another
  formula without $\neg$ operator.
\end{enumerate}
The remaining proof is routine and is omitted here.
\end{proof}

Not surprisingly $\WBS$ is not preserved by $\interleave$, which can
be shown by using the same arguments as in
Counterexample~\ref{cex:weak bisimulation composition branching}.

\subsection{Taxonomy for weak bisimulation}
\begin{figure}[t]
  \begin{center}
    \begin{picture}(40,40)(-20,-0)
    \gasset{Nw=16,Nh=6,Nmr=1}
    \node(WBSB)(-20,20){$\WBSB$}
    \node(P)(-20,40){$\PCTL_{\backslash\X}$}
    \node(WEBS)(20,20){$\WBS$}
    \node(PS)(20,40){$\PCTL^{*}_{\backslash\X}$}
    \gasset{Nw=10,Nh=10,Nmr=5}
    \node(BSP)(0,0){$\bBSP$}
    \drawedge[exo=2,sxo=2](BSP,WBSB){}
    \drawedge[exo=2,sxo=2](BSP,WEBS){}
    \drawedge[AHnb=1,ATnb=1](P,WBSB){}
    \drawedge[AHnb=1,ATnb=1](PS,WEBS){}
    \drawedge[eyo=2,syo=2](PS,P){}
    \drawedge[eyo=2,syo=2](WEBS,WBSB){}
    \gasset{ELdistC=y,ELdist=0}
    \drawedge[ELside=r,exo=-2,sxo=-2](WBSB,BSP){$\setminus$}
    \drawedge[ELside=r,exo=-2,sxo=-2](WEBS,BSP){$\setminus$}
    \drawedge[ELside=r,eyo=-2,syo=-2](P,PS){$\setminus$}
    \drawedge[ELside=r,eyo=-2,syo=-2](WBSB,WEBS){$\setminus$}
    \end{picture}
  \end{center}
\caption{\label{fig:summary of relation weak}
 Relationship of different equivalences in weak scenario.}
\end{figure}
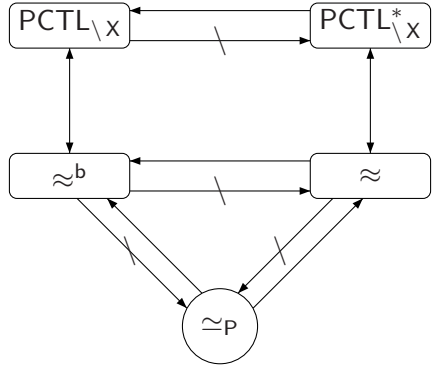

As in the strong case, we summarize the equivalence relations in the
weak scenario in Fig.~\ref{fig:summary of relation weak}, where all the
denotations have the same meanings as Fig.~\ref{fig:summary of
  relation}. Compared to Fig.~\ref{fig:summary of relation},
Fig.~\ref{fig:summary of relation weak} is much simpler because the
step-indexed bisimulations are absent. As in the strong case, we
do not consider the standard definition of weak bisimulation~\cite{SegalaL95}
which is a strict subset of $\bBSP$ and can be defined by replacing
$\bTRANP{}{}$ with $\bTRAN{}{}$ in Definition~\ref{def:branching
  bisimulation}. Again, not surprisingly, all the relations shown in
Fig.~\ref{fig:summary of relation weak} coincide with the weak
bisimulation defined in~\cite{BaierKHW05} over the {\sf DTMC}s setting.

\section{Simulations}\label{sec:simulation}
In Section~\ref{sec:strong} and \ref{sec:weak} we have discussed
bisimulations and their characterizations.  Two states $s$ and
$r$ are bisimilar iff $s$ can mimic stepwise all the transitions of
$r$ and vice versa.  In this section we consider simulation relations
that only require one direction mimicking. Simulations are preorders on
the states, which have been used widely for verification
purposes~\cite{milner1989communication,Jonsson91Simulation,Henzinger95Computing,SegalaL95,BaierKHW05}.
If $r$ simulates $s$, then $s$ can be seen as a correct
implementation of $r$. Since $r$ is more abstract and contains less
details, it is more preferable to be analysed, moreover some properties satisfied
by $r$ are also preserved by $s$.

We shall discuss the characterization of simulations w.r.t. the safe
fragments of $\PCTL$ and $\PCTL^{*}$.  First let us introduce the safe
fragment of $\PCTLS$, denoted by $\PCTLS_{\MI{safe}}$, which is a
fragment of $\PCTLS$ without negative operators except for the atomic
propositions, and is defined by the following syntax:
\begin{align*}
\phi &::= a \mid\neg a\mid \phi_1\land\phi_2\mid \phi_1\lor\phi_2\mid\MC{P}_{\leq q}(\psi)\\
\psi &::=\phi\mid\psi_1\land\psi_2\mid\psi_1\lor\psi_2\mid\X\psi\mid\psi_1\U\psi_2
\end{align*}
where $a\in\AP$ and $q\in[0,1]$. Accordingly, the safe fragment of $\PCTL$, denoted by $\PCTL_{\MI{safe}}$,
is a sub logic of $\PCTLS_{\MI{safe}}$ where the path formulas are constrained to be the following form:
$$\psi::=\X\phi\mid\phi_1\U\phi_2 \mid\phi_1\U^{\le n}\phi_2.$$

We write $s~\Si_{\PCTLS_{\MI{safe}}}~r$ iff $r\models\phi$ implies that $s\models\phi$ for any $\phi$ of $\PCTLS_{\MI{safe}}$, and similarly for other sub-logics.

Below we recall the notion of \emph{weight
  functions}~\cite{jonsson1991specification}, and then use them to
define strong probabilistic simulation relations~\cite{SegalaL95}:
\begin{defi}\label{def:weight function}
Let $\MC{R}=S\times S$ be a relation over $S$. A weight function for $\mu$ and $\nu$ with respect to $\MC{R}$ is a function $\Delta:S\times S\mapsto[0,1]$ such that:
\begin{iteMize}{$\bullet$}
\item $\Delta(s,r)>0$ implies that $s~\MC{R}~r$,
\item $\mu(s)=\sum_{r\in S}\Delta(s,r)$ for any $s\in S$,
\item $\nu(r)=\sum_{s\in S}\Delta(s,r)$ for any $r\in S$.
\end{iteMize}
We write $\mu~\DSI~\nu$ iff there exists a weight function for $\mu$ and $\nu$ with respect to $\MC{R}$.
\end{defi}

\begin{defi}\label{def:strong probabilistic simulation}
A relation $\MC{R}\subseteq S\times S$ is a strong
probabilistic simulation iff $s~\MC{R}~r$ implies that $L(s)=L(r)$ and for each $\TRAN{s}{\mu}$, there exists a combined transition
$\TRANP{r}{\mu'}$ such that $\mu~\DSI~\mu'$.

We write $s~\SP~r$ whenever there is a strong
probabilistic simulation $\MC{R}$ such that $s~\MC{R}~r$.
\end{defi}

It was shown in~\cite{SegalaL95} that $\DSI$ is a congruence,
i.e. $s~\SP~r$ implies that $\PAR{s}{t}~\SP~\PAR{r}{t}$ for any $t$.
But not surprisingly, it turns out that the strong probabilistic simulation is too fine
w.r.t $\Si_{\PCTL_{\MI{safe}}}$ and $\Si_{\PCTLS_{\MI{safe}}}$ which can be seen from Example~\ref{ex:counterexample}.
Similarly we have an analogue of Theorem~\ref{thm:PCLT and Probabilistic bisimulation} in the simulation scenario,
where we only consider the safe fragment of the logics,
thus the subscription $\MI{safe}$ is often omitted for readability.
\begin{thm}\label{thm:PCLT and Probabilistic simulation}\hfill
\begin{enumerate}[\em(1)]
\item $\SEPCTL$, $\SEPCTLS$, $\SEPCTLM$, $\iSEPCTLM{i}$, $\SEPCTLSM$, $\iSEPCTLSM{i}$, and $\SP$ are preorders for any $i\geq 1$.
\item $\SP~\subset~\SEPCTLS~\subset~\SEPCTL$.
\item $\SEPCTLSM~\subset~\SEPCTLM$.
\item $\iSEPCTLSM{1}~=~\iSEPCTLM{1}$.
\item $\iSEPCTLSM{i}~\subset~\iSEPCTLM{i}$ for any $i>1$.
\item $\SEPCTL~\subset~\SEPCTLM~\subset~\iSEPCTLM{i+1}~\subset~\iSEPCTLM{i}$ for all $i\geq 0$.
\item $\SEPCTLS~\subset~\SEPCTLSM~\subset~\iSEPCTLSM{i+1}~\subset~\iSEPCTLSM{i}$ for all $i\geq 0$.
\end{enumerate}
\end{thm}
\begin{proof}
For Clause (1) we only prove that $\SEPCTL$ is a preorder since the others are similar. The reflexivity is trivial as $s~\SEPCTL~s$ for any $s$. Suppose that $s~\SEPCTL~t$ and $t~\SEPCTL~r$, then we need to prove that $s~\SEPCTL~r$ in order to show the transitivity. According to the definition of $\SEPCTL$, we need to prove that $r\models\phi$ implies $s\models\phi$ for any $\phi$. Suppose that $r\models\phi$ for some $\phi$, then $t\models\phi$ because of $t~\SEPCTL~r$, moreover since $s~\SEPCTL~t$, hence $s\models\phi$ which completes the proof.

The proof of Clause (2) can be found in~\cite{SegalaL95}. 

In order to prove Clause (4), we can follow
the same reasoning as in Theorem~\ref{thm:PCLT and Probabilistic bisimulation},
by showing that the safe fragments of $\PCTL^{-}_{1}$ and $\PCTL^{*-}_{1}$  coincide. Again the syntax of path formulas of safe $\PCTL^{*-}_{1}$
can be rewritten as:
$$\psi ::= \phi \mid \X\phi\mid\psi_1\land\psi_2\mid\psi_1\lor\psi_2,$$
therefore only the following cases need to be considered:
$\MC{P}_{\le q}(\phi)$, $\MC{P}_{\le q}(\X\phi_1\land\X\phi_2)$, 
$\MC{P}_{\le q}(\X\phi_1\land\phi_2)$, $\MC{P}_{\le q}(\X\phi_1\lor\X\phi_2)$,
$\MC{P}_{\le q}(\X\phi_1\lor\phi_2)$, all of which can be transformed to a formula
in safe $\PCTL^{-}_1$. For instance $s\models\MC{P}_{\le q}(\X\phi_1\lor\phi_2)$
iff $s\models\phi_2\lor s\models\MC{P}_{\le q}(\X\phi_1)$. The remaining proof
is similar and omitted here.

The proofs of all the other clauses are trivial.
\end{proof}

\subsection{Strong \texorpdfstring{$i$}{i}-depth branching simulation}\label{sec:strong i brancing simulation}
Following Section~\ref{sec:i depth bisimulation branching} we can define strong $i$-depth branching simulation which can be characterized by $\iSEPCTLM{i}$. Let $s~\iBSi{0}~r$ iff $L(s)=L(r)$, then
\begin{defi}\label{def:index strong branching simulation}
A relation $\MC{R}\subseteq S\times S$ is a
strong $i$-depth branching simulation with $i\geq1$ iff $s~\MC{R}~r$ implies that $s~\iBSi{i-1}~r$ and for any $\MC{R}$ downward closed sets $C, C'$, and any scheduler $\sigma$, there exists $\sigma'$ such that $\MEASURE_{\sigma',r}(C,C',i,r)\geq\MEASURE_{\sigma,s}(C,C',i,s)$.

We write $s~\iBSi{i}~r$ whenever there is a strong $i$-depth branching simulation $\MC{R}$ such that $s~\MC{R}~r$. The strong branching simulation $\iBSi{}$ is defined as $\iBSi{}\ =\ \cap_{i\geq 0} \iBSi{i}$.
\end{defi}

Below we show properties similar to Lemma~\ref{lem:i equivalence relation branching} for strong $i$-depth branching simulation.
\begin{lem}\label{lem:i preorder relation branching}\hfill
\begin{enumerate}[\em(1)]
\item $\iBSi{}$ and $\iBSi{i}$ are preorders for any $i\geq0$.
\item $\iBSi{j}~\subseteq~\iBSi{i}$ provided that $0\leq i \leq j$.
\item There exists $i\geq 0$ such that $\iBSi{j}~=~\iBSi{k}$ for any $j,k\geq i$.
\end{enumerate}
\end{lem}
\begin{proof}
We consider the first clause. The reflexivity is trivial, we only prove the
  transitivity. Suppose that $s_1~\iBSi{i}~s_2$ and
  $s_2~\iBSi{i}~s_3$, we need to prove that $s_1~\iBSi{i}~s_3$. By
  Definition~\ref{def:index strong branching simulation} there exists
  strong simulation $\MC{R}_1$ and $\MC{R}_2$ such that
  $s_1~\MC{R}_1~s_2$ and $s_2~\MC{R}_2~s_3$. 
 Moreover since $s~\iBSi{i}~s$ for any $s$,  reflexive relations $\MC{R}_1$ and $\MC{R}_2$  
  always exist. Let
  $\MC{R}=\MC{R}_1\circ\MC{R}_2=\{(s_1,s_3)\mid\exists
  s_2.(s_1~\MC{R}_1~s_2\land s_2~\MC{R}_2~s_3)\}$, it is enough to
  prove that $\MC{R}$ is strong $i$-depth branching simulation. Due to
  the reflexivity, any $\MC{R}$ downward closed set $C$ is also
  $\MC{R}_1$ and $\MC{R}_2$ downward closed. Therefore for any
  $\MC{R}$ downward closed sets $C,C'$ and a scheduler $\sigma$, 
  there exists $\sigma'$ such that
  $\MEASURE_{\sigma',s_2}(C,C',i,s_2)\geq\MEASURE_{\sigma,s_1}(C,C',i,s_1)$
  according to Definition~\ref{def:index strong branching
    simulation}. Similarly, there exists $\sigma''$ such that
  $\MEASURE_{\sigma'',s_3}(C,C',i,s_3)\geq\MEASURE_{\sigma',s_2}(C,C',i,s_2)\geq\MEASURE_{\sigma,s_1}(C,C',i,s_1)$,
  and $\MC{R}$ is indeed a strong $i$-depth branching simulation. This
  completes the proof.

The second clause follows directly  from Definition~\ref{def:index strong branching simulation}.

For the third clause, note there are only finitely many states, thus in the worst case each state
is only able to simulate itself, and there always exists $i$ such that $\iBS{j}$ is stable for all 
$j \ge i$.
\end{proof}

Our strong $i$-depth branching simulation coincides with $\iSEPCTLM{i}$ for each $i$, therefore $\SEPCTL$ is equivalent to $\iBSi{}$ as shown by the following theorem.
\begin{thm}\label{thm:characterization strong branching simulation}
$\iSEPCTLM{i}~=~\iBSi{i}$ for any $i\geq 1$, and moreover $\SEPCTL~=~\iBSi{}$.
\end{thm}
\begin{proof}
We first prove that $\iSEPCTLM{i}$ implies $\iBSi{i}$. Let $\MC{R}=\{(s,r)\mid s~\iSEPCTLM{i}~r\}$ and $s~\MC{R}~r$, we need to prove that for any $\MC{R}$ downward closed sets $C,C'$ and scheduler $\sigma$ of $s$, there exists $\sigma'$ of $r$ such that $\MEASURE_{\sigma',r}(C,C',i,r)\geq\MEASURE_{\sigma,s}(C,C',i,s)$. Note that $\MI{Sat}(\phi)$ is a $\MC{R}$ downward closed set for any $\phi$. Since the states space is finite, for each $\MC{R}$ downward closed set $C$, there exists $\phi_C$ such that $\MI{Sat}(\phi_C)=C$. Assume that there exists $\MC{R}$ downward closed sets $C,C'$ and $\sigma$ such that $\MEASURE_{\sigma',r}(C,C',i,r)<\MEASURE_{\sigma,s}(C,C',i,s)$ for all schedulers $\sigma'$ of $r$. Then there exists $\MEASURE_{\sigma',r}(C,C',i,r)\le q<\MEASURE_{\sigma,s}(C,C',i,s)$ such that $r\models\MC{P}_{\leq q}(\psi)$ but $s\not\models\MC{P}_{\leq q}(\psi)$ where $\psi=\phi_C\U^{\leq i}\phi_{C'}$, this contradicts the assumption that $s~\iSEPCTLM{i}~r$. Therefore $\MC{R}$ is a strong $i$-depth branching simulation.

In order to prove that $\iBSi{i}$ implies $\iSEPCTLM{i}$, we need to prove that whenever $s~\iBSi{i}~r$ and $r\models\phi$, we also have $s\models\phi$. We prove by structural induction on $\phi$, and only consider the case when $\phi=\MC{P}_{\leq q}(\phi_1\U^{\leq i}\phi_2)$ since all the others are trivial. By induction hypothesis $\MI{Sat}(\phi_1)$ and $\MI{Sat}(\phi_2)$ are $\iBSi{i}$ downward closed, therefore if $r\models\MC{P}_{\leq q}(\phi_1\U^{\leq i}\phi_2)$, but $s\not\models\MC{P}_{\leq q}(\phi_1\U^{\leq i}\phi_2)$, then there exists a scheduler $\sigma$ of $s$ such that there does not exist $\sigma'$ such that $\MEASURE_{\sigma',r}(\MI{Sat}(\phi_1),\MI{Sat}(\phi_2),i,r)\geq\MEASURE_{\sigma,s}(\MI{Sat}(\phi_1),\MI{Sat}(\phi_2),i,s)$ which contradicts the assumption that $s~\iBSi{i}~r$.
\end{proof}

In Counterexample~\ref{cex:i compositional branching} we have shown
the $\iBSB{i}$ is not compositional for $i>1$, using the same
arguments we can show that $\iBSi{i}$ is not compositional either for
$i>1$, thus we have:
\begin{lem}\label{lemma:i compositional branching simulation}
$s~\iBSi{1}~r$ implies that $\PAR{s}{t}~\iBSi{1}~\PAR{r}{t}$ for any $t$, while $\iBSi{i}$ with $i>1$ is not compositional in general.
\end{lem}
\begin{proof}
Let $\MC{R}=\{(\PAR{s}{t},\PAR{r}{t})\mid s~\iBSi{1}~r\}$, it is enough to show that $\MC{R}$ is a strong 1-depth simulation.
Let $C'$ be a $\iBSi{1}$ downward closed set, then $\{\PAR{s'}{t}\mid s'\in C'\}$ is $\MC{R}$ downward closed, the following
proof is similar with the proof of Lemma~\ref{lem:1-composition}.

Note that Counterexample~\ref{cex:i compositional branching} also applies here, thus $\iBSi{i}$ is not necessarily compositional when $i>1$.
\end{proof}

\begin{rem}
The safe fragment of $\PCTL$ we adopt in this paper is slightly different from~\cite{BaierKHW05} where two new operators $\widetilde{\X}$ and $\widetilde{\U}$ are introduced, called weak next and until respectively, and $\MC{P}_{\leq q}(\psi)$ is replaced by $\MC{P}_{\geq q}(\psi)$. The semantics of $\widetilde{\X}$ and $\widetilde{\U}$ are defined as follows where $\ABS{\omega}$ denotes the length of $\omega$:
\begin{align*}
\omega\models\widetilde{\X}\phi &\text{ iff } (\ABS{\omega}<1\lor\omega[i]\models\phi)\\
\omega\models\phi_1\widetilde{\U}\phi_2&\text{ iff } (\omega\models\phi_1\U\phi_2\lor\forall i\leq\ABS{\omega}.\omega[i]\models\phi_1)
\end{align*}
Similarly we can also define the weak counterpart of the bounded until
$\widetilde{\U}^{\leq n}$. Due to the duality between $\X$, $\U^{\leq n}$,
$\U$ and their weak counterparts, these two variants of safe $\PCTL$
are essentially equivalent, and we refer to~\cite{BaierKHW05} for detailed
discussions.

Let $\PCTL_{\MI{live}}$ denote the liveness fragment of $\PCTL$ in~\cite{BaierKHW05} which is the same as $\PCTL_{\MI{safe}}$ except that $\MC{P}_{\leq q}(\psi)$ is replaced with $\MC{P}_{\geq q}(\psi)$. We say $s~\Si_{\PCTL_{\MI{live}}}~r$ iff $s\models\phi$ implies $r\models\phi$ for any state formula of $\PCTL_{\MI{live}}$. Even though it has been shown in~\cite{BaierKHW05} that $\Si_{\PCTL_{\MI{safe}}}$ and $\Si_{\PCTL_{\MI{live}}}$ are equivalent for $\DTMC$s, the result is not true for $\PA$s. Refer to the following example.
\begin{exa}
  Consider the two states $s_0$ and $r_0$ shown in
  Fig.~\ref{fig:liveness}, where we assume that all the states have
  different labels except that $L(s_0)=L(r_0)$. It is easy to check
  that $s_0~\SP~r_0$, thus $s_0~\Si_{\PCTL_{\MI{safe}}}~r_0$ according
  to Clause (2) of Theorem~\ref{thm:PCLT and Probabilistic
    simulation}, but we have
  $s_0~\not\Si_{\PCTL_{\MI{live}}}~r_0$. Let $\phi=\MC{P}_{\geq
    1}(L(s_0)\U L(s_1))$ which is a valid state formula of
  $\PCTL_{\MI{live}}$, it is obvious that $s_0\models\phi$, but
  $r_0\not\models\phi$ since the minimal probability of $r_0$ reaching
  state $s_1$ is equal to 0 i.e. by choosing the transition to $s_2$.
\end{exa}

\begin{figure}[t]
  \begin{center}
    \begin{picture}(80,20)(0,-0)
    \gasset{Nadjust=n,Nw=6, Nh=6,Nmr=3}
    \node(AA)(0,20){$s_0$}
    \node(AB)(0,0){$s_1$}
    \node(BA)(60,20){$r_0$}
    \node(BB)(50,0){$s_1$}
    \node(BC)(70,0){$s_2$}
    \drawedge(AA,AB){1}
    \drawedge[ELside=r](BA,BB){1}
    \drawedge(BA,BC){1}
    \drawloop[loopangle=180](AB){1}
    \drawloop[loopangle=180](BB){1}
    \drawloop[loopangle=0](BC){1}
    \end{picture}
  \end{center}
\caption{\label{fig:liveness}An example illustrating $s_0~\not\Si_{\PCTL_{\MI{live}}}~r_0$.}
\end{figure}
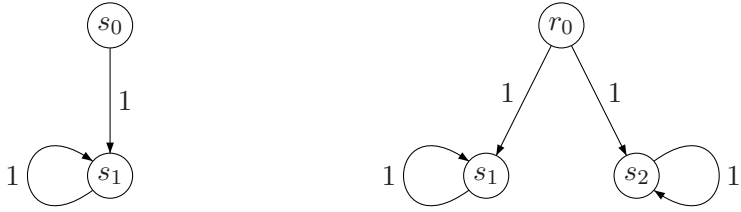
\end{rem}

\subsection{Strong \texorpdfstring{$i$}{i}-depth simulation}
In this section we introduce strong $i$-depth simulation which can be characterized by $\iSEPCTLSM{i}$. Below follows the definition of strong $i$-depth simulation where $\iSi{0}~=~\iBSi{0}$.
\begin{defi}\label{def:index strong simulation}
A relation $\MC{R}\subseteq S\times S$ is a strong $i$-depth simulation with $i\geq1$ iff $s~\MC{R}~r$ implies that $s~\iSi{i-1}~r$ and for any $\widetilde{\Omega}\subseteq (\DOWNWARD{\MC{R}}{})^+$ with $l(\tilde{\Omega})=i$ and any scheduler $\sigma$, there exists $\sigma'$ such that $\MEASURE_{\sigma',r}(C_{\widetilde{\Omega}})\geq\MEASURE_{\sigma,s}(C_{\widetilde{\Omega}})$.

We write $s~\iSi{i}~r$ whenever there is a strong $i$-depth simulation $\MC{R}$ such that $s~\MC{R}~r$. The strong simulation $\iSi{}$ is defined as $\iSi{}\ =\ \cap_{i\geq 0}\iSi{i}$.
\end{defi}

Below we show properties similar to Lemma~\ref{lem:i equivalence relation} for strong $i$-depth simulations.
\begin{lem}\label{lem:i preorder relation}\hfill
\begin{enumerate}[\em(1)]
\item $\iSi{}$ and $\iSi{i}$ are preorders for any $i\geq0$.
\item $\iSi{j}~\subseteq~\iSi{i}$ provided that $0\leq i \leq j$.
\item There exists $i\geq 0$ such that $\iSi{j}~=~\iSi{k}$ for any $j,k\geq i$.
\end{enumerate}
\end{lem}
\proof\hfill
\begin{enumerate}[(1)]
\item This clause can be proved in a similar way as Clause (1) of Lemma~\ref{lem:i preorder relation branching}.
\item According to Definition~\ref{def:index strong simulation}, as
  $i$ is increasing, $\iSi{i}$ is getting finer. 
\item The proof is based on the fact that the number of states is
  finite, with the similar argument as in Clause (3) of Lemma~\ref{lem:i preorder relation branching}.\qed
\end{enumerate}

\noindent Our strong $i$-depth simulation coincides with
$\iSEPCTLSM{i}$ for each $i$, therefore $\SEPCTLS$ is equivalent to
$\iSi{}$ as shown by the following theorem.
\begin{thm}\label{thm:characterization strong simulation}
$\iSEPCTLSM{i}~=~\iSi{i}$ for any $i\geq 1$, and moreover $\SEPCTLS~=~\iSi{}$.
\end{thm}

\proof
  We first prove that $s~\iSEPCTLSM{i}~r$ implies $s~\iSi{i}~r$ for
  any $s$ and $r$. Let $\MC{R}=\{(s,r)\mid s~\iSEPCTLSM{i}~r\}$ and
  $s~\MC{R}~r$, we need to show that for any
  $\widetilde{\Omega}\subseteq (\DOWNWARD{\MC{R}}{})^+$ with
  $l(\widetilde{\Omega})\leq i$ and scheduler $\sigma$, there exists a
  scheduler $\sigma'$ such that
  $\MEASURE_{\sigma',r}(C_{\widetilde{\Omega}})\geq\MEASURE_{\sigma,s}(C_{\widetilde{\Omega}})$. By
  the definition of $\MC{R}$, there exists a formula $\phi_C$ such
  that $\mathit{Sat}(\phi_C)=C$ where $C$ is an $\MC{R}$ downward closed
  set. Suppose $\Omega=C_0C_1\ldots C_j$ with $j\leq i$, then
$$\psi_{\Omega}=\phi_{C_0}\land\X(\phi_{C_1}\land\ldots\land\X(\phi_{C_{j-1}}\land\X\phi_{C_j})\ldots)
$$
can be used to characterize $\Omega$, that is, $\mathit{Sat}(\psi_{\Omega})=C_{\Omega}$. Let $\psi=\mathop{\lor}\limits_{\Omega\in\widetilde{\Omega}}\psi_{\Omega}$, then $\mathit{Sat}(\psi)=C_{\widetilde{\Omega}}$. We proceed by contradiction. Suppose that there does not exist $\sigma'$ such that $\MEASURE_{\sigma',r}(C_{\widetilde{\Omega}})\geq\MEASURE_{\sigma,s}(C_{\widetilde{\Omega}})$, then there exists $q$ such that $r\models\MC{P}_{\leq q}(\psi)$, but $s\not\models\MC{P}_{\leq q}(\psi)$ which contradicts the assumption that $s~\iSEPCTLSM{i}~r$, so there exists a scheduler $\sigma'$ such that $\MEASURE_{\sigma',r}(C_{\widetilde{\Omega}})\geq q =\MEASURE_{\sigma,s}(C_{\widetilde{\Omega}})$.

The proof of $\iSi{i}~\subseteq~\iSEPCTLSM{i}$ is by structural induction on the syntax of state formula $\phi$ and path formula $\psi$ of safe $\PCTL^{*-}_i$, that is, we need to prove the following two results simultaneously.
\begin{enumerate}[(1)]
\item $r\models\phi$ implies $s\models\phi$ for any state formula $\phi$ provided that $s~\iSi{i}~r$.
\item $\omega_2\models\psi$ implies $\omega_1\models\psi$ for any path formula $\psi$ provided that $\omega_1~\iSi{i}~\omega_2$.
\end{enumerate}

We only consider $\phi=\MC{P}_{\leq q}(\psi)$ such that $\DEPTH(\psi)\leq i$ here. Suppose that 
$r\models\phi$, i.e. $\forall\sigma.\MEASURE_{\sigma,r}(\{\omega\mid\omega\models\psi\})\leq
q$, we need to show that $s\models\phi$. We proceed by contradiction, and assume that 
$s\not\models\phi$, i.e. there exists $\sigma$ such that 
$\MEASURE_{\sigma,s}(\{\omega\mid\omega\models\psi\})>q$. 
By induction hypothesis $\{\omega\mid\omega\models\psi\}$ is $\iSi{i}$ 
downward closed. Since $\DEPTH(\psi)\leq i$, there exists 
$\widetilde{\Omega}$ such that $l(\widetilde{\Omega})\leq i$ and 
$C_{\widetilde{\Omega}}= \{\omega\mid\omega\models\psi\}$.  
Since $r\models\phi$, there does not exists $\sigma'$ such that 
$\MEASURE_{\sigma',r}(C_{\widetilde{\Omega}})\geq\MEASURE_{\sigma,s}(C_{\widetilde{\Omega}})=q$, which 
contradicts the assumption that $s~\iSi{i}~r$, thus it holds that $s\models\phi$.\qed

Similarly, we can show that $\iSi{i}$ is not compositional either for
$i>1$, thus we have:
\begin{lem}\label{lemma:i compositional simulation}
$s~\iSi{1}~r$ implies that $\PAR{s}{t}~\iSi{1}~\PAR{r}{t}$ for any $t$, while $\iSi{i}$ with $i>1$ is not compositional in general.
\end{lem}
\begin{proof}
According to Theorem~\ref{thm:characterization strong branching simulation} and \ref{thm:characterization strong simulation}, and Clause (4) of Theorem~\ref{thm:PCLT and Probabilistic simulation}, $\iBSi{1}~=~\iSi{1}$, thus the result is straightforward according to Lemma~\ref{lemma:i compositional branching simulation}.
\end{proof}

\subsection{Weak simulation}
Given the results for weak bisimulation from Section~\ref{sec:weak}, the characterization of weak simulation is straightforward.
Let us first introduce the definition of weak probabilistic simulation by Segala and Lynch~\cite{SegalaL95} as follows:
\begin{defi}\label{def:branching simulation}
A relation $\MC{R}\subseteq S\times S$ is a weak probabilistic simulation iff $s~\MC{R}~r$ implies that $L(s)=L(r)$ and for each $\TRAN{s}{\mu}$, there exists $\bTRANP{r}{\mu'}$ such that $\mu~\DSI~\mu'$.

We write $s~\bSiP~r$ whenever there is a weak probabilistic simulation $\MC{R}$ such that $s~\MC{R}~r$.
\end{defi}

\noindent From~\cite{SegalaL95} we know that $\bSiP$ is compositional, but it is too fine for $\SEPCTLWN$ as well as $\SEPCTLSWN$, therefore along the line of weak bisimulation, we have similar results for weak simulation. Below follows the definition of  branching simulation.

\begin{defi}\label{def:weak simulation branching}
A relation $\MC{R}\subseteq S\times S$ is a branching simulation iff $s~\MC{R}~r$ implies that $L(s)=L(r)$ and for any $\MC{R}$ downward closed sets $C,C'$ and any scheduler $\sigma$, there exists $\sigma'$ such that $\MEASURE_{\sigma',r}(C,C',r)\geq\MEASURE_{\sigma,s}(C,C',s)$.

We write $s~\WBSi~r$ whenever there is a branching simulation $\MC{R}$ such that $s~\MC{R}~r$.
\end{defi}

\noindent Due to Counterexample~\ref{cex:weak bisimulation composition
  branching}, $\WBSi$ is not compositional, but it coincides with
$\SEPCTLWN$ as shown by the following theorem.
\begin{thm}\label{thm:weak simulation and PCTL branching}
$\WBSi$ is a preorder, and $\WBSi~=~\SEPCTLWN$.
\end{thm}
\begin{proof}
The proof of the first statement is along the
  same line as the proof of Clause (1) of Lemma~\ref{lem:i preorder relation branching}.
For the second statement, 
 we prove that $s~\SEPCTLWN~r$ implies $s~\WBSi~r$ for
  any $s$ and $r$. Let $\MC{R}=\{(s,r)\mid
  s~\SEPCTLWN~r\}$ and $s~\MC{R}~r$, we need to
  prove that for any $\MC{R}$ downward closed sets $C,C'$ and
  scheduler $\sigma$, there exists a scheduler $\sigma'$ such that
  $\MEASURE_{\sigma',r}(C,C',r)\geq\MEASURE_{\sigma,s}(C,C',s)$. Let $\phi_C$ be a formula such that
  $\mathit{Sat}(\phi_C)=C$ where $C$ is a $\MC{R}$ downward closed
  set. We proceed by contradiction. Suppose that there does not exist
  $\sigma'$ such that
  $\MEASURE_{\sigma',r}(C,C',r)\geq\MEASURE_{\sigma,s}(C,C',s)$, then
  there exists $q$ such that $r\models\MC{P}_{\leq q}(\psi)$ where
  $\psi=\phi_C\U\phi_{C'}$, but $s\not\models\MC{P}_{\leq q}(\psi)$,
  which contradicts the assumption that $s~\SEPCTLWN~r$. Therefore
  there must exist a scheduler $\sigma'$ such that
  $\MEASURE_{\sigma',r}(C,C',r)\geq\MEASURE_{\sigma,s}(C,C',s)$.

  The proof of $\WBSi~\subseteq~\SEPCTLWN$ is by structural induction
  on the syntax of state formula $\phi$ and path formula $\psi$ of
  safe $\PCTL_{\backslash\X}$, that is, we need to prove the following
  two results simultaneously.
\begin{enumerate}[(1)]
\item $r\models\phi$ implies $s\models\phi$ for any state formula $\phi$ provided that $s~\WBSi~r$.
\item $\omega_2\models\psi$ implies that $\omega_1\models\psi$ for any path formula $\psi$ provided that $\omega_1~\WBSi~\omega_2$.
\end{enumerate}
We only consider $\phi=\MC{P}_{\leq q}(\psi)$ where
$\psi=\phi_1\U\phi_2$ since the other cases are similar. Suppose that
$r\models\phi$, we need to prove that $s\models\phi$. We proceed by
contradiction, and assume that $s\not\models\phi$, then there exists
$\sigma$ such that
$\MEASURE_{\sigma,s}(\{\omega\mid\omega\models\psi\})>q$. By induction hypothesis
$\mathit{Sat}(\phi_1)$ and $\mathit{Sat}(\phi_2)$ are $\WBSi$ downward
closed, thus
$\MEASURE_{\sigma,s}(\mathit{Sat}(\phi_1),\mathit{Sat}(\phi_2),s)=\MEASURE_{\sigma,s}(\{\omega\mid\omega\models\psi\})>q$. Since
$r\models\phi$, there does not exist $\sigma'$ such that
$\MEASURE_{\sigma,r}(\mathit{Sat}(\phi_1),\mathit{Sat}(\phi_2),r)\geq
\MEASURE_{\sigma,s}(\mathit{Sat}(\phi_1),\mathit{Sat}(\phi_2),s)$
which contradicts the assumption that $s~\WBSi~r$, thus
$s\models\phi$, and $s~\SEPCTLWN~r$.
\end{proof}

The weak simulation equivalent to $\SEPCTLSWN$ can also be obtained in a straightforward way by adapting Definition~\ref{def:weak bisimulation}.
\begin{defi}\label{def:weak simulation}
A relation $\MC{R}\subseteq S\times S$ is a
weak simulation iff $s~\MC{R}~r$ implies that $L(s)=L(r)$ and for any $\widetilde{\Omega}\subseteq(\DOWNWARD{\MC{R}}{})^+$ and any scheduler $\sigma$, there exists $\sigma'$ such that $\MEASURE_{\sigma',r}(C_{\widetilde{\Omega}_{\mathit{st}}})\geq\MEASURE_{\sigma,s}(C_{\widetilde{\Omega}_{\mathit{st}}})$.

We write $s~\WSi~r$ whenever there is a weak simulation $\MC{R}$ such that $s~\MC{R}~r$.
\end{defi}

Again $\WSi$ is not compositional, but it coincides with $\SEPCTLSWN$,
therefore we have the following theorem.
\begin{thm}\label{thm:weak simulation and PCTL}
$\WSi$ is a preorder, and $\WSi~=~\SEPCTLSWN$.
\end{thm}
\begin{proof}
  The reflexivity of $\WSi$ is trivial. We prove the transitivity of
  $\WSi$. Suppose that $s~\WSi~r$ and $r~\WSi~t$, then for any
  $\widetilde{\Omega}\subseteq(\DOWNWARD{\WSi}{})^+$ and scheduler
  $\sigma$, there exists $\sigma'$ such that
  $\MEASURE_{\sigma',r}(C_{\widetilde{\Omega}_{\MI{st}}})\geq\MEASURE_{\sigma,s}(C_{\widetilde{\Omega}_{\MI{st}}})$. Since
  we also have $r~\WBSi~t$, so there exists $\sigma''$ such that
  $\MEASURE_{\sigma'',t}(C_{\widetilde{\Omega}_{\MI{st}}})\geq\MEASURE_{\sigma',r}(C_{\widetilde{\Omega}_{\MI{st}}})\geq\MEASURE_{\sigma,s}(C_{\widetilde{\Omega}_{\MI{st}}})$. This
  proves the transitivity of $\WSi$.

  For the second statement let $\MC{R}=\{(s,r)\mid s~\SEPCTLSWN~r\}$ and
  $s~\MC{R}~r$, we need to prove that for any
  $\widetilde{\Omega}\subseteq(\DOWNWARD{\MC{R}}{})^+$ and scheduler
  $\sigma$, there exists a scheduler $\sigma'$ such that
  $\MEASURE_{\sigma',r}(C_{\widetilde{\Omega}_{\mathit{st}}})\geq\MEASURE_{\sigma,s}(C_{\widetilde{\Omega}_{\mathit{st}}})$. By
  induction hypothesis $C_{\widetilde{\Omega}_{\mathit{st}}}$ is
  $\MC{R}$ downward closed, thus there exists $\psi$ such that
  $\MI{Sat}(\psi)=C_{\widetilde{\Omega}_{\mathit{st}}}$. We proceed by
  contradiction. Suppose that there does not exist $\sigma'$ such that
  $\MEASURE_{\sigma',r}(C_{\widetilde{\Omega}_{\mathit{st}}})\geq\MEASURE_{\sigma,s}(C_{\widetilde{\Omega}_{\mathit{st}}})$,
  then there exists $q$ such that $r\models\MC{P}_{\leq q}(\psi)$, but
  $s\not\models\MC{P}_{\leq q}(\psi)$, which contradicts the
  assumption that $s~\SEPCTLSWN~r$. Therefore there must exist a
  scheduler $\sigma'$ such that
  $\MEASURE_{\sigma',r}(C_{\widetilde{\Omega}_{\mathit{st}}})\geq\MEASURE_{\sigma,s}(C_{\widetilde{\Omega}_{\mathit{st}}})$.

  The proof of $\WSi~\subseteq~\SEPCTLSWN$ is by structural induction
  on the syntax of state formula $\phi$ and path formula $\psi$ of
  safe $\PCTL_{\backslash\X}^{*}$, that is, we need to prove the
  following two results simultaneously.
\begin{enumerate}[(1)]
\item $r\models\phi$ implies $s\models\phi$ for any state formula $\phi$ provided that $s~\WSi~r$.
\item $\omega_2\models\psi$ implies that $\omega_1\models\psi$ for any path formula $\psi$ provided that $\omega_1~\WSi~\omega_2$.
\end{enumerate}
We only consider $\phi=\MC{P}_{\geq q}(\psi)$ since the other cases
are similar. Suppose that $r\models\phi$, we need to prove that
$s\models\phi$. We prove by contradiction, and assume that
$s\not\models\phi$, then there exists $\sigma$ such that
$\MEASURE_{\sigma,s}(\{\omega\mid\omega\models\psi\})<q$. By induction
hypothesis $\{\omega\mid\omega\models\psi\}$ is $\WSi$ downward
closed, thus there exists $\widetilde{\Omega}_{\MI{st}}$ such that
$C_{\widetilde{\Omega}_{\MI{st}}}=\{\omega\mid\omega\models\psi\}$. Since
$r\models\phi$, there does not exist $\sigma'$ such that
$\MEASURE_{\sigma,r}(C_{\widetilde{\Omega}_{\MI{st}}})\geq
\MEASURE_{\sigma,s}(C_{\widetilde{\Omega}_{\MI{st}}})=q$ which contradicts
the assumption that $s~\WSi~r$, thus $s\models\phi$, and
$s~\SEPCTLSWN~r$.
\end{proof}

\subsection{Simulation kernels and summary}
Let $\MC{R}^{-1}$ denote the reverse of $\MC{R}$, then
$\MC{R}\cap\MC{R}^{-1}$ is the simulation kernel. In this section we
will show the relation between the simulation kernels and their
correspondent bisimulations. Not surprisingly, the simulation kernels
are coarser than the bisimulations as shown in the following lemma:
\begin{lem}\hfill
\begin{enumerate}[\em(1)]
\item $\iBSB{i}~\subset~(\iBSi{i}\cap(\iBSi{i})^{-1})$.
\item $\iBS{i}~\subset~(\iSi{i}\cap\iSi{i}^{-1})$.
\item $\WBSB~\subset~(\WBSi\cap(\WBSi)^{-1})$.
\item $\WBS~\subset~(\WSi\cap\WSi^{-1})$.
\end{enumerate}
\end{lem}
\begin{proof}
  We only prove the first clause here, since the others are quite
  similar.  The inclusion
  $\iBSB{i}~\subseteq~\iBSi{i}\cap(\iBSi{i})^{-1}$ is straightforward.
  To show that the inclusion is strict, it is enough to give a
  counterexample. Suppose we have three states $s_1$, $s_2$, and $s_3$
  such that $s_1~\iBSi{i}~s_2~\iBSi{i}~s_3$ but
  $s_3~\not\iBSi{i}~s_2~\not\iBSi{i}~s_1$. Let $s$ and $r$ be two
  states such that $L(s)=L(r)$. In addition $s$ has three transitions:
  $\TRAN{s}{\DIRAC{s_1}}$, $\TRAN{s}{\DIRAC{s_2}}$,
  $\TRAN{s}{\DIRAC{s_3}}$, and $r$ only has two transitions:
  $\TRAN{r}{\DIRAC{s_1}}$, $\TRAN{r}{\DIRAC{s_3}}$. Then it should be
  easy to check that $s~\iBSi{i}~r$ and $r~\iBSi{i}~s$, the only
  non-trivial case is when $\TRAN{s}{\DIRAC{s_2}}$. Since
  $s_2~\iBSi{i}~s_3$, thus there exists $\TRAN{r}{\DIRAC{s_3}}$ such
  that $\DIRAC{s_2}~\DSI[\iBSi{i}]~\DIRAC{s_3}$. But obviously
  $s~\not\iBSB{i}~r$, since the transition $\TRAN{s}{\DIRAC{s_2}}$
  cannot be simulated by any transition of $r$.
\end{proof}

The relationship of all the preorders is similar as for bisimulations,
which we have summarized in Fig.~\ref{fig:summary of relation} and
\ref{fig:summary of relation weak} respectively.  We could draw
similar figures for the preorders by replacing all the equivalence
relations by their correspondent preorders, but we omit them here for
lack of space.

\section{Countable states}\label{sec:countable}
Until now we have only considered $\PA$s with finitely many states. In this section we will show that these results also apply for $\PA$s with countable states.
Assume $S$ is a countable set of states.
We adopt the method used in~\cite{DesharnaisGJP10} to deal with strong branching bisimulation. First we recall some standard notations from topology theory. Given a metric space $(S,d)$ where $d$ is a metric, a sequence $\{s_i\mid i\geq 0\}$ converges to $s$ iff for any $\epsilon>0$, there exists $n$ such that $d(s_m,s)<\epsilon$ for any $m\geq n$. A metric space $(S,d)$ is compact if every infinite sequence has a convergent subsequence to an element in $S$.

Below follows the definition of metric over distributions: Refer to~\cite{DesharnaisGJP10} for more details.
\begin{defi}\label{def:metric}
Given two distributions $\mu,\nu\in\mathit{Dist}(S)$, the metric $d$ is defined by $d(\mu,\nu)=\SUP_{C\subseteq S}\ABS{\mu(C)-\nu(C)}$.
\end{defi}

Since the metric is defined over distributions, we need to adapt the
definition of $\MEASURE_{\sigma,s}(C,C',n,s)$ in the following way:
$s\boundTRAN{n}{C}\mu$ iff either i) $\mu=\DIRAC{s}$, or ii)
$s\TRAN{}\nu$ such that $$\sum\limits_{\forall
  r\in\SUPP(\nu).r\boundTRAN{n-1}{C}\nu_r}\nu(r)\cdot\nu_r=\mu.$$ Obviously for each $\sigma,C,C'$, and $n$, there exists
$s\boundTRAN{n}{C}\mu$ such that
$\mu(C')=\MEASURE_{\sigma,s}(C,C',n,s)$.

Now we can define the compactness of $\PA$s as in~\cite{DesharnaisGJP10} with a slight difference.
\begin{defi}\label{def:compactness}
Given a $\PA$ $\MC{P}$, $\MC{P}$ is $i$-compact iff the metric space $(\{\mu\mid s\boundTRAN{i}{C}\mu\},d)$ is compact for each $s\in S$ and $\iBSB{i}$ closed set $C$.
\end{defi}

As mentioned in~\cite{DesharnaisGJP10,schaefer1999topological}, the convex closure does not change the compactness, thus we can extend $\boundTRAN{n}{C}$ to allow combined transitions in a standard way without changing anything, but for simplicity we omit this. A $\PA$ is \emph{compact} iff it is $i$-compact for any $i\geq 1$.

We introduce the definition of \emph{capacity} as follows.
\begin{defi}\label{def:capacity}
  Given a set of states $S$ and a $\sigma$-algebra $\MC{B}$, a
  capacity on $\MC{B}$ is a function $\MI{Cap}:\MC{B}\rightarrow
  R^{\ge 0}$ such that:
\begin{enumerate}[(1)]
\item $\MI{Cap}(\emptyset)=0$,
\item whenever $C_1\subseteq C_2$ with $C_1,C_2\in\MC{B}$, then
  $\MI{Cap}(C_1)\leq\MI{Cap}(C_2)$,
\item whenever there exists $C_1\subseteq C_2\subseteq\ldots$ such that $\cup_{i\geq 1}C_i=C$, or $C_1\supseteq C_2\supseteq\ldots$ such that $\cap_{i\geq 1}C_i=C$, then $\lim_{i\rightarrow\infty}\MI{Cap}(C_i)=\MI{Cap}(C)$.
\end{enumerate}
A capacity $\MI{Cap}$ is \emph{sub-additive} iff $\MI{Cap}(C_1\cup C_2)\leq\MI{Cap}(C_1) + \MI{Cap}(C_2)$ for any $C_1,C_2\in\MC{B}$.
\end{defi}

Different from~\cite{DesharnaisGJP10}, the value of $\MEASURE_{\sigma,s}(C,C',n,s)$ depends on both $C$ and $C'$. Let $\MEASUREONE_{s,n}^{C}(C')=\SUP_{\sigma}\MEASURE_{\sigma,s}(C,C',n,s)$ and $\MEASURETWO_{s,n}^{C'}(C)=\SUP_{\sigma}\MEASURE_{\sigma,s}(C,C',n,s)$ i.e. given a $C'$, $\MEASUREONE_{s,n}^C$ will return the maximum probability from $s$ to $C'$ in at most $n$ steps via only states in $C$, similar for $\MEASURETWO_{s,n}^{C'}$. The following lemma shows that both $\MEASUREONE_{s,n}^{C}$ and $\MEASURETWO_{s,n}^{C'}$ are sub-additive capacities.
\begin{lem}\label{lem:capacity}
$\MEASUREONE_{s,n}^{C}$ and $\MEASURETWO_{s,n}^{C'}$ are sub-additive capacities on $\MC{B}$ where $\MC{B}$ is the $\sigma$-algebra only containing $\iBSB{n}$ closed sets.
\end{lem}
\begin{proof}
Refer to the proof of Lemma 5.2 in~\cite{DesharnaisGJP10}.
\end{proof}

Now we can show that the following results are still valid as long as the given $\PA$ is compact
even if it contains infinitely countable states.
\begin{thm}\label{thm:infinite i branching}
Given a compact $\PA$,
\begin{enumerate}[\em(1)]
\item $\iBSB{n}~=~\iEPCTLM{n}$,
\item there exists $n\geq 0$ such that $\iBSB{n}~=~\EPCTL$.
\end{enumerate}
\end{thm}
\proof\hfill
\begin{enumerate}[(1)]
\item $\iBSB{n}~=~\iEPCTLM{n}$:\\
The proof of $\iBSB{n}~\subseteq~\iEPCTLM{n}$ is similar with
  the proof of Theorem~\ref{thm:relation of equivalence branching},
  and is omitted here. We prove that $\iEPCTLM{n}~\subseteq~\iBSB{n}$
  in the sequel following the proof of Theorem 6.10
  in~\cite{DesharnaisGJP10}. Let $\MC{R}=\{(s,r)\mid
  s~\iEPCTLM{n}~r\}$, we need to prove that $\MC{R}$ is a strong
  $n$-depth branching bisimulation. In order to do so, we need to
  prove that for any $(s,r)\in\MC{R}$, if
  $\MEASURE_{\sigma,s}(C,C',n,s)>0$ for some $\sigma$, there exists
  $\sigma'$ such that
  $\MEASURE_{\sigma',r}(C,C',n,r)\geq\MEASURE_{\sigma,s}(C,C',n,s)$
  and vice versa. This is equivalent to say that
  $\SUP_{\sigma}\MEASURE_{\sigma,s}(C,C',n,s)=\SUP_{\sigma}\MEASURE_{\sigma,r}(C,C',n,s)$,
  i.e. $\MEASUREONE_{s,n}^{C}(C')=\MEASUREONE_{r,n}^{C}(C')$ (or
  equivalently $\MEASURETWO_{s,n}^{C'}(C)=\MEASURETWO_{r,n}^{C'}(C)$)
for each $\MC{R}$ closed sets $C$ and $C'$. Since
  both $C$ and $C'$ may be countable union of equivalence classes
  where each equivalence class can only be characterized by countable
  many formulas, therefore we have
  $C=\cup_{i=1}^{\infty}(\cap_{j=1}^{\infty}C_{i,j})$ and
  $C'=\cup_{i=1}^{\infty}(\cap_{j=1}^{\infty}C'_{i,j})$ where
  $\cap_{j=1}^{\infty}C_{i,j}$ corresponds to the
  $i$-th equivalence class in $C$, and $C_{i,j}$ corresponds  to
  the set of states determining by the $j$-th formula
  satisfied by $i$-th equivalence class, similar for
  $\cap_{j=1}^{\infty}C'_{i,j}$ and $C'_{i,j}$. Let $B_k =
  \cap_{j=1}^{\infty}(\cup_{i=1}^{k}C_{i,j})$,
  $A_k^l=\cap_{j=1}^{l}(\cup_{i=1}^{k}C_{i,j})$, and $B'_k =
  \cap_{j=1}^{\infty}(\cup_{i=1}^{k}C'_{i,j})$,
  $A_k^{'l}=\cap_{j=1}^{l}(\cup_{i=1}^{k}C'_{i,j})$. It is easy to see
  that $B_k$ and $B'_k$ are increasing sequences of $\MC{R}$ closed
  sets such that $\cup_{k=1}^{\infty}B_k=C$, and
  $\cup_{k=1}^{\infty}B'_k=C'$, while $A_k^l$ and $A_k^{'l}$ are
  decreasing sequences of $\MC{R}$ closed sets such that
  $\cap_{l=1}^{\infty}A_k^l=B_k$ and
  $\cap_{l=1}^{\infty}A_k^{'l}=B'_k$. Both $A_k^l$ and $A_k^{'l}$ only
  contain conjunction and disjunction of finite formulas, thus can be
  described by $\PCTL_n^{-}$, that is, let
  $\phi=\land_{j=1}^l(\lor_{i=1}^k\phi_{C_{i,j}})$ and
  $\phi'=\land_{j=1}^l(\lor_{i=1}^k\phi_{C'_{i,j}})$ where
  $\phi_{C_{i,j}}$ denotes the $j$-th formula satisfied by the $i$-th
  equivalence class in $C$, similarly for
  $\phi_{C'_{i,j}}$. Obviously, we have $\mathit{Sat}(\phi)=A^l_k$ and
  $\mathit{Sat}(\phi')=A^{'l}_k$.

  Assume that
  $q=\MEASUREONE_{r,n}^{A_k^l}(A_k^{'l})<\MEASUREONE_{s,n}^{A_k^l}(A_k^{'l})=p$,
  then it holds that $r\models\MC{P}_{\leq q}(\phi\U^{\leq n}\phi')$,
  but $s\not\models\MC{P}_{\leq q}(\phi\U^{\leq n}\phi')$, which
  contradicts the fact that $s~\iEPCTLM{n}~r$. Therefore
  $\MEASUREONE_{s,n}^{A_k^l}(A_k^{'l})=\MEASUREONE_{r,n}^{A_k^l}(A_k^{'l})$
  for each $l$ and $k$.  By Definition~\ref{def:capacity} and
  Lemma~\ref{lem:capacity}, we know that
  $\MEASUREONE_{s,n}^{A_k^l}(C')=\MEASUREONE_{r,n}^{A_k^l}(C')$ for
  each $l$ and $k$. Note that
  $\MEASUREONE_{s,n}^{A_k^l}(C')=\MEASURETWO_{s,n}^{C'}(A_k^l)$, thus
  $\MEASURETWO_{s,n}^{C'}(A_k^l)=\MEASURETWO_{r,n}^{C'}(A_k^l)$ for
  each $l$ and $k$, again by Definition~\ref{def:capacity} and
  Lemma~\ref{lem:capacity}, we conclude that
  $\MEASURETWO_{s,n}^{C'}(C)=\MEASURETWO_{r,n}^{C'}(C)$.
  \item  $\exists n\geq 0.\iBSB{n}~=~\EPCTL$:\\
  Suppose that $\EPCTL~\subset~\iBSB{n}$ for any $n\geq 0$ which
  means that there exist $s$ and $r$ such that $s~\iBSB{n}~r$ for any
  $n\geq 0$, but $s~\nEPCTL~r$. As a result there exists $C,C'$ and
  $\sigma$ such that
  $\lim_{i\rightarrow\infty}\MEASURE_{\sigma,s}(C,C',i,s)>0$, but
  there does not exist $\sigma'$ such that
  $\lim_{i\rightarrow\infty}\MEASURE_{\sigma',r}(C,C',i,r)\geq\lim_{i\rightarrow\infty}\MEASURE_{\sigma,s}(C,C',i,s)$. In
  other words,
  $\lim_{i\rightarrow\infty}\MEASURE_{\sigma',r}(C,C',i,r)<\lim_{i\rightarrow\infty}\MEASURE_{\sigma,s}(C,C',i,s)$
  for any $\sigma'$ which indicates that there exists $n\geq 0$ such
  that $\MEASURE_{\sigma',r}(C,C',n,r)<\MEASURE_{\sigma,s}(C,C',n,s)$
  for any $\sigma'$, therefore $s~\inEPCTLM{i}~r$ which contradicts
  our assumption.\qed
\end{enumerate}

\noindent In a similar way we can extend the results of this section to strong
bisimulations and weak bisimulations: We skip the proofs here. For
the simulations, we need to do more work, since there may be
uncountably many downward closed sets.  For a relation $\MC{R}$ over
$S$, let $\equiv_{\MC{R}}$ denote the largest equivalence relation
contained in the reflexive and transitive closure of
$(\MC{R}\cup\MI{ID})$. The following lemma states that a downward
closed set can be expressed as a union of equivalence classes:

\begin{lem}\label{lem:downward equivalence}
Let $\MC{R}~\subseteq~S\times S$ be a relation, and $C\subseteq S$ be a $\MC{R}$ downward closed set, then $C$ is a union of equivalence classes of $\equiv_{\MC{R}}$.
\end{lem}
The above lemma is a slight generalization of Lemma 5.1
in~\cite{HermannsPSWZ11} with only two differences: i) we consider
downward closed sets instead of upward closed sets, ii) we do not
require $\MC{R}$ to be a preorder, but these do not change the proof
there.

Given a $\MC{R}$ downward closed set $C$, we say $C$ is \emph{finitely generated} if there exists a finite set of equivalence classes of $\{ C_i\in S/\equiv_{\MC{R}}\}_{i\in I}$ such that $C=\cup_{i\in I}C_i$. Since the set of the equivalence classes in $S/\equiv_{\MC{R}}$ is countable, thus the set of finitely generated $\MC{R}$ downward closed set is also countable~\cite{HermannsPSWZ11}. The following lemma shows an alternative definition of $\iBSi{i}$ in Definition~\ref{def:index strong branching simulation} where we only focus on finitely generated downward closed sets:
\begin{lem}\label{lem:finitely generated}
A relation $\MC{R}\subseteq S\times S$ is a
strong $i$-depth branching simulation with $i\geq1$ iff $s~\MC{R}~r$ implies that $s~\iBSi{i-1}~r$ and for any finitely generated $\MC{R}$ downward closed sets $C, C'$, and any scheduler $\sigma$, there exists $\sigma'$ such that $\MEASURE_{\sigma',r}(C,C',i,r)\geq\MEASURE_{\sigma,s}(C,C',i,s)$.

We write $s~\iBSi{i}~r$ whenever there is a strong $i$-depth branching simulation $\MC{R}$ such that $s~\MC{R}~r$.
\end{lem}
\begin{proof}
  The proof is similar as the proof of Lemma 5.2
  in~\cite{HermannsPSWZ11}. Let $(\iBSi{i})'$ denote the new
  definition, we need to prove that $s~\iBSi{i}~r$ iff
  $s~(\iBSi{i})'~r$. Since finitely generated $\MC{R}$ downward closed
  sets are special cases of $\MC{R}$ downward closed sets, therefore
  $s~\iBSi{i}~r$ implies $s~(\iBSi{i})'~r$. We prove that
  $s~(\iBSi{i})'~r$ implies $s~\iBSi{i}~r$ by contradiction. Suppose
  that for any finitely generated $\MC{R}$ downward closed sets $C,C'$
  and $\sigma$, there exists $\sigma'$ such that
  $\MEASURE_{\sigma',r}(C,C',i,r)\geq\MEASURE_{\sigma,s}(C,C',i,s)$, but
  there exists $\MC{R}$ downward closed sets $C,C'$ and $\sigma$ such
  that $\MEASURE_{\sigma',r}(C,C',i,r)<\MEASURE_{\sigma,s}(C,C',i,s)$ for
  any $\sigma'$. Let $\sigma$ be a scheduler such that
  $\MEASURE_{\sigma',r}(C,C',i,r)<\MEASURE_{\sigma,s}(C,C',i,s)$ for any
  $\sigma'$ and
  $\epsilon=\MEASURE_{\sigma,s}(C,C',i,s)-\MEASURE_{\sigma',r}(C,C',i,r)>0$. According
  to Lemma~\ref{lem:downward equivalence}, there exists sets of
  equivalences classes: $\{C_j\in S/\equiv_{\MC{R}}\}_{j\in J}$ and
  $\{C_k\in S/\equiv_{\MC{R}}\}_{k\in K}$ such that $C=\cup_{j\in
    J}C_i$ and $C'=\cup_{k\in K}C_k$ where $J,K$ are (infinite) sets
  of indexes. Define two sequences of finitely generated $\MC{R}$
  downward closed sets: $\{C_{\leq j}=\cup_{j'\in J\land j'\leq
    j}C_{j'}\}_{ j\in J}$, $\{C_{\leq k}=\cup_{k'\in K\land k'\leq k}C_{k'}\}_{k\in K}.$ Obviously both $\MEASURE_{\sigma,s}(C,C_{\leq k},i,s)$ and
  $\MEASURE_{\sigma,s}(C_{\leq j},C',i,s)$ are monotone, non-decreasing
  and converge to $\MEASURE_{\sigma,s}(C,C',i,s)$ for any $C$ and
  $C'$. Therefore there exists $j\in J$ and $k\in K$ such that
$$\MEASURE_{\sigma,s}(C_{\leq j},C',i,s)>\MEASURE_{\sigma,s}(C,C',i,s) - \frac{\epsilon}{4}, \text{ and }$$
$$\MEASURE_{\sigma,s}(C_{\leq j},C_{\leq k},i,s)>\MEASURE_{\sigma,s}(C_{\leq j},C',i,s) - \frac{\epsilon}{4}.$$
This implies
$$\MEASURE_{\sigma,s}(C_{\leq j},C_{\leq k},i,s) > \MEASURE_{\sigma,s}(C,C',i,s) - \frac{\epsilon}{2}$$$$=\MEASURE_{\sigma',r}(C,C',i,r) + \frac{\epsilon}{2}>\MEASURE_{\sigma',r}(C,C',i,r)\geq \MEASURE_{\sigma,s}(C_{\leq j},C_{\leq k},i,s),$$
which contradicts the assumption.
\end{proof}

By Lemma~\ref{lem:finitely generated} it is enough to consider all the finitely generated $\iBSi{i}$ downward closed sets in Definition~\ref{def:compactness} which is countable. The extension of Theorem~\ref{thm:characterization strong branching simulation} to the countable state space is then routine i.e. we should define the capacity as in Definition~\ref{def:capacity}, and then show that finite formulas are enough to characterize $\MEASURE_{\sigma',s}(C,C',i,s)$ even if $C$ and $C'$ are countable infinite. Moreover the definitions of other variants of simulations in Section~\ref{sec:simulation} can be adopted to only consider finitely generated downward closed sets too, thus their logic characterizations can also be extended to countable states.

\section{The coarsest congruent bisimulations and simulations}\label{sec:congruent}
Before we have shown that $\BSP$ is a congruence but cannot be
characterized by $\EPCTL$ since it is too fine. On the
other hand,  $\iBSB{}$ can be characterized by
$\EPCTL$, but it is not a congruence in general. This indicates that 
$\EPCTL$ is not a congruence. Therefore a natural question
one may ask is what is the largest subset of $\EPCTL$ which is
congruent. The following theorem shows that $\BSP$ is the coarsest
congruence relation in $\EPCTL$ provided that the given $\PA$ is compact.
\begin{thm}\label{thm:coarsest}
Given a compact $\PA$, $\BSP$ is the coarsest congruence relation in $\EPCTL$.
\end{thm}
\begin{proof}
  We proceed by contradiction. Suppose that there exists a congruence
  $\simeq~\subset~\EPCTL$. Suppose that there exists
  $s$ and $r$ such that $s~\simeq~r$ but $s~\nBSP~r$. According to
  Definition~\ref{def:strong probabilistic bisimulation} there exists
  $s\TRAN{}\mu$ such that there does not exist $r\TRANP{}\nu$ with
  $\mu~\BSP~\nu$. The idea is to show that there always exists $t$
  such that $\PAR{s}{t}~\not\EPCTL~\PAR{r}{t}$ in this case, then it
  is enough to give a formula $\phi$ such that
  $\PAR{r}{t}\models\phi$, but $\PAR{s}{t}\not\models\phi$.

Let $\SUPP(\mu)=\{s_1,s_2,\ldots\}$ and $\mu(s_i)=a_i$ with $i\geq 1$,
where we assume that $s_i(i\geq 1)$ belong to different
  equivalence classes for simplicity. Without loss of generality we
assume that there exists $s\TRAN{}\mu$ such that for any two
(combined) transitions of $r$: $r\TRANP{}\nu_1$ and $r\TRANP{}\nu_2$,
there does not exist $0\leq w_1,w_2\leq 1$ such that $w_1 + w_2 = 1$
and $\mu~\BSP~(w_1\cdot\nu_1 + w_2\cdot\nu_2)$ (every combined
transition of $r$ can be seen as a combined transition of two other
combined transitions of $r$). Let $\nu_1(s_i)=b_i$ and
$\nu_2(s_i)=c_i$ in the following, then there must exist $i\neq j\geq
1$ such that there is no $0\leq w_1,w_2\leq 1$ with $w_1\cdot b_i
+ w_2\cdot c_i=a_i$, $w_1\cdot b_j + w_2\cdot c_j=a_j$, and
$w_1+w_2=1$, otherwise we will have $\mu~\BSP~(w_1\cdot\nu_1 +
w_2\cdot\nu_2)$ which contradicts the assumption.

Most of the cases are simple, for instance if $a_i>b_i,c_i$, $r$
will evolve into $s_i$ with probability less than $a_i$ which is not
the case for $s$, thus $s~\nEPCTL~r$ which contradicts the assumption.
We only consider in detail the case when $c_i> b_i$, $b_j> c_j$, $a_i\in(b_i,c_i)$ and $a_j\in(c_j,b_j)$.
Suppose that $\frac{b_j-a_j}{a_i-b_i}=\frac{a_j-c_j}{c_i-a_i}$,
which implies that $\frac{a_i-b_i}{c_i-a_i}=\frac{b_j-a_j}{a_j-c_j}$. Let
$w_1=\frac{1}{k+1}$ and $w_2=\frac{k}{k+1}$ where $k=\frac{a_i-b_i}{c_i-a_i}$,
then it holds that $w_1\cdot b_i + w_2\cdot c_i=a_i$ and $w_1\cdot b_j + w_2\cdot c_j=a_j$, which contradicts the assumption. Therefore we have either $\frac{b_j-a_j}{a_i-b_i}<\frac{a_j-c_j}{c_i-a_i}$
or $\frac{b_j-a_j}{a_i-b_i}>\frac{a_j-c_j}{c_i-a_i}$. We only consider the case
when $\frac{b_j-a_j}{a_i-b_i}>\frac{a_j-c_j}{c_i-a_i}$, since the other case
is similar. Let $\rho_1$ and $\rho_2$ be
two variables with values in $[0,1]$ such that
$$\frac{b_j-a_j}{a_i-b_i}\cdot\rho_2 > \rho_1 > \frac{a_j-c_j}{c_i-a_i}\cdot\rho_2,$$
then we can see that:
$$
a_i\cdot\rho_1 + a_j\cdot\rho_2 < b_i\cdot\rho_1 + b_j\cdot\rho_2,$$
$$a_i\cdot\rho_1 + a_j\cdot\rho_2 < c_i\cdot\rho_1 + c_j\cdot\rho_2.
$$
In other words, there exists $\rho_1$ and $\rho_2$ such that
$a_i\cdot\rho_1 + a_j\cdot\rho_2$ is smaller than $b_i\cdot\rho_1 + b_j\cdot\rho_2$ and $c_i\cdot\rho_1 + c_j\cdot\rho_2$.

Let $t$ be a state such that it can only evolve into $t_1$ with
probability $\rho_1$ and $t_2$ with probability $\rho_2$ where $\rho_1+\rho_2=1$ and
$\rho_1\in(\frac{a_j-c_j}{c_i-a_i}\cdot\rho_2,\frac{b_j-a_j}{a_i-b_i}\cdot\rho_2)$;
obviously such $t$ always exists. Assume that all the states have
distinct labels except for $s$ and $r$, moreover let
\[\psi=((L(s\interleave t)\lor L(s_i\interleave
t)\lor(L(s_j\interleave t)))\U^{\leq 2} (L(s_i\interleave t_1)\lor
L(s_j\interleave t_2))),\] it is not hard to see that the minimum
probability of the paths of $\PAR{s}{t}$ satisfying $\psi$ is
$a_i\cdot\rho_1 + a_j\cdot\rho_2$ i.e. when $\PAR{s}{t}$ first
performs the transition $s\TRAN{}\mu$ of $s$ and then performs the
transition $t\TRAN{}\{\rho_1:t_1,\rho_2:t_2\}$ of $t$. Let
$r\TRANP{}\nu=w_1\cdot\nu_1+w_2\cdot\nu_2$ 
be a transition for some $w_1$ and $w_2$ such that after performing it,
the probability of the set of paths of $\PAR{r}{t}$ satisfying $\psi$ is
minimal. It holds:
\begin{align*}
\nu(s_i)\cdot\rho_1+\nu(s_j)\cdot\rho_2 &=(w_1\cdot b_i+w_2\cdot c_i)\cdot\rho_1 + (w_1\cdot b_j + w_2\cdot c_j)\cdot\rho_2\\
& = w_1\cdot(b_i\cdot\rho_1 + b_j\cdot\rho_2) + w_2\cdot(c_i\cdot\rho_1 + c_j\cdot\rho_2)\\
& > a_i\cdot\rho_1 + a_j\cdot\rho_2,
\end{align*}
therefore we have $\PAR{r}{t}\models\MC{P}_{\geq q}(\psi)$
but $\PAR{s}{t}\not\models\MC{P}_{\geq q}(\psi)$ where
$q=\nu(s_i)\cdot\rho_1+\nu(s_j)\cdot\rho_2$. In other words
$\PAR{s}{t}~\nEPCTL~\PAR{r}{t}$, as a result
$\PAR{s}{t}~\not\simeq~\PAR{r}{t}$, so $\simeq$ is not a congruence.

When all the states do not have distinct labels,
we can always construct formulas to distinguish them,
since the $\PA$ is compact and these states are in different equivalence classes by assumption.
The subsequent proof is then similar. This completes our proof.
\end{proof}

Theorem~\ref{thm:coarsest} can be extended to identify the coarsest congruent weak bisimulation in $\EPCTLWN$, and the coarsest congruent strong and weak simulations in $\SEPCTL$ and $\SEPCTLWN$ respectively.
\begin{thm}\label{thm:coarsest 1}\hfill
\begin{enumerate}[\em(1)]
\item $\bBSP$ is the coarsest congruence relation in $\EPCTLWN$,
\item $\SP$ is the coarsest congruent preorder in $\SEPCTL$,
\item $\bSiP$ is the coarsest congruent preorder in $\SEPCTLWN$.
\end{enumerate}
\end{thm}
\begin{proof}
The proof is similar to the proof of Theorem~\ref{thm:coarsest} and we only sketch the proof of Clause (2) here. According to Lemma 5.2 in~\cite{HermannsPSWZ11}, $\mu~\MC{R}~\nu$ iff for each finitely generated $\MC{R}$ downward closed set $C$, $\mu(C)\leq\nu(C)$ where $\MC{R}$ is a preorder. In order to prove that $\SP$ is the coarsest congruent preorder in $\SEPCTL$, we need to show that for any relation $\preceq$ such that $\SP~\subset~\preceq~\subset~\SEPCTL$, it holds that $\preceq$ is not congruent, i.e. there exist $s$, $r$, and $t$ such that $s~\preceq~r$, but $\PAR{s}{t}~\not\preceq~\PAR{r}{t}$. First assume that $\preceq$ is a congruence, and we then prove by contradiction as in Theorem~\ref{thm:coarsest} and show that if $s~\preceq~r$ and $s~\not\SP~r$, there exists $t$ such that $\PAR{s}{t}~\not\SEPCTL~\PAR{r}{t}$, thus $\PAR{s}{t}~\not\preceq~\PAR{r}{t}$ which contradicts the assumption that $\preceq$ is a congruence. Since $s~\not\SP~r$, then there exists $s\TRAN{}\mu$ such that there does not exist $r\TRANP{}\nu$ with $\mu~\DSI[\SP]~\nu$. With the same argument as in Theorem~\ref{thm:coarsest} and Lemma 5.2 in~\cite{HermannsPSWZ11}, there exist $t$ and $\psi$ such that $\PAR{r}{t}\models\MC{P}_{\geq q}(\psi)$ but $\PAR{s}{t}\not\models\MC{P}_{\geq q}(\psi)$ i.e. $\PAR{s}{t}~\not\SEPCTL~\PAR{r}{t}$, thus $\preceq$ is not congruent.
\end{proof}

\section{Related work}\label{sec:related}
For Markov chains, i.e., deterministic $\PA$s, the
logic $\PCTL$ characterizes bisimulations, and $\PCTL$ without $\X$ operator
characterizes weak bisimulations~\cite{HanssonJ90,BaierKHW05}.  As pointed out
in~\cite{SegalaL95}, probabilistic bisimulation  is sound, but not complete for $\PCTL$ over $\PA$s.
In the literature, various extensions of the Hennessy-Milner logic~\cite{HennessyM85} are considered for characterizing
bisimulations. Larsen and Skou~\cite{larsen1991bisimulation} considered such an
extension of Hennessy-Milner logic, which characterizes bisimulation for
\emph{reactive probabilistic processes}~\cite{larsen1991bisimulation}. Similar results are further studied for labelled Markov
processes~\cite{prakash-book,DesharnaisGJP10} (with continuous state space).  For $\PA$s, Jonsson
\emph{et al.}~\cite{Jonsson} considered a two-sorted logic in the
Hennessy-Milner style to characterize strong bisimulations. In~\cite{HermannsPSWZ11}, the results are 
also extended to characterize simulations.

\begin{figure}[!t]
  \centering
    \begin{picture}(140, 40)(0,0)
    \gasset{Nw=6,Nh=6,Nmr=3}
    \node(SAA)(0,0){$s_1$}
    \node(SBA)(10,0){$s_2$}
    \node(SCA)(20,0){$s_3$}
    \node(SAB)(30,0){$s_1$}
    \node(SBB)(40,0){$s_2$}
    \node(SCB)(50,0){$s_3$}
    \node(RAA)(60,0){$s_1$}
    \node(RBA)(70,0){$s_2$}
    \node(RCA)(80,0){$s_3$}
    \node(RAB)(90,0){$s_1$}
    \node(RBB)(100,0){$s_2$}
    \node(RCB)(110,0){$s_3$}
    \node(RAC)(120,0){$s_1$}
    \node(RBC)(130,0){$s_2$}
    \node(RCC)(140,0){$s_3$}
    \node(S)(25,40){$s$}
    \node(SMA)(13,20){$s_{\mu_1}$}
    \node(SMC)(37,20){$s_{\mu_3}$}
    \node(R)(100,40){$r$}
    \node(RMA)(80,20){$s_{\mu_1}$}
    \node(RMB)(100,20){$s_{\mu_2}$}
    \node(RMC)(120,20){$s_{\mu_3}$}
    \drawedge(S,SMA){}
    \drawedge(S,SMC){}
    \drawedge(R,RMA){}
    \drawedge(R,RMB){}
    \drawedge(R,RMC){}
    \gasset{ELdistC=y,ELdist=0}
    \drawedge[ELpos=60,ELside=r](SMA,SAA){{\tiny \colorbox{white}{0.3}}}
    \drawedge[ELpos=60,ELside=r](SMA,SBA){{\tiny \colorbox{white}{0.3}}}
    \drawedge[ELpos=60,ELside=r](SMA,SCA){{\tiny \colorbox{white}{0.4}}}
    \drawedge[ELpos=60,ELside=r](SMC,SAB){{\tiny \colorbox{white}{0.5}}}
    \drawedge[ELpos=60,ELside=l](SMC,SBB){{\tiny \colorbox{white}{0.4}}}
    \drawedge[ELpos=60,ELside=l](SMC,SCB){{\tiny \colorbox{white}{0.1}}}
    \drawedge[ELpos=60,ELside=r](RMA,RAA){{\tiny \colorbox{white}{0.3}}}
    \drawedge[ELpos=60,ELside=r](RMA,RBA){{\tiny \colorbox{white}{0.3}}}
    \drawedge[ELpos=60,ELside=r](RMA,RCA){{\tiny \colorbox{white}{0.4}}}
    \drawedge[ELpos=60,ELside=r](RMB,RAB){{\tiny \colorbox{white}{0.4}}}
    \drawedge[ELpos=60,ELside=r](RMB,RBB){{\tiny \colorbox{white}{0.3}}}
    \drawedge[ELpos=60,ELside=l](RMB,RCB){{\tiny \colorbox{white}{0.3}}}
    \drawedge[ELpos=60,ELside=r](RMC,RAC){{\tiny \colorbox{white}{0.5}}}
    \drawedge[ELpos=60,ELside=l](RMC,RBC){{\tiny \colorbox{white}{0.4}}}
    \drawedge[ELpos=60,ELside=l](RMC,RCC){{\tiny \colorbox{white}{0.1}}}
    \end{picture}
  \caption{\label{fig:altercounterexample} Alternating automata.}
\end{figure}
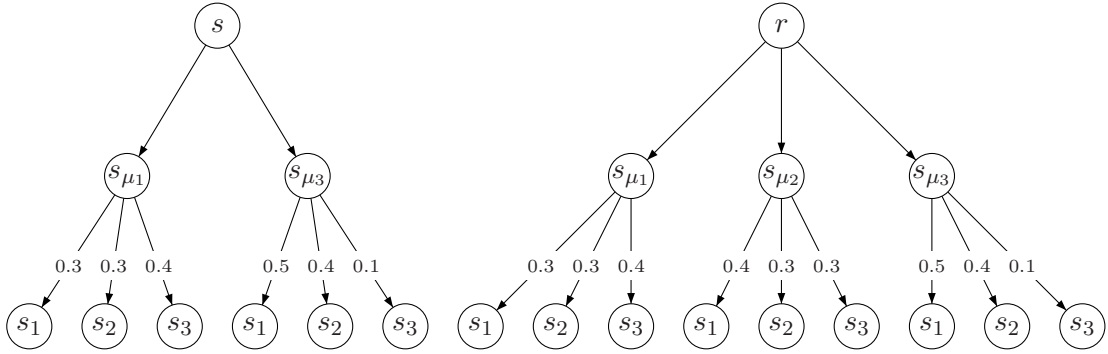

Weak bisimulation was first defined in the context of $\PA$s by
Segala and Lynch~\cite{SegalaL95}, and then formulated for alternating models by
Philippou \emph{et al.}~\cite{PhilippouLS00}.
The seemingly very related work is by Desharnais et
al.~\cite{DesharnaisGJP10}, where it is shown that $\PCTL^{*}$ is sound and
complete with respect to weak bisimulation for \emph{alternating
automata}. The key difference is that the model they have considered is not
the same as $\PA$s considered in this paper. Briefly,
in alternating automata, states are either non-deterministic like in
transition systems, or stochastic like in discrete-time Markov chains.
As discussed in~\cite{SegalaT05}, a $\PA$ can be
transformed to an alternating automaton by replacing each transition
$\TRANA{s}{}{ \mu}$ by two consecutive transitions $\TRANA{s}{}{s'}$
and $\TRANA{s'}{}{\mu}$ where $s'$ is the new inserted
state. Surprisingly, for alternating automata, Desharnais et al. have
shown that weak bisimulation -- defined in the standard manner --
characterizes $\PCTL^{*}$ formulas. The following example illustrates why it
works in that setting, but fails for $\PA$s.
\begin{exa}\label{ex:alternative PA}
  Refer to Fig.~\ref{fig:counterexample}, we need to add three
  additional states $s_{\mu_1},s_{\mu_2},$ and $s_{\mu_3}$ in order to
  transform $s$ and $r$ to states in an alternating automaton. 
  The resulting
  automaton is shown in Fig.~\ref{fig:altercounterexample}.  Suppose
  that $s_1,s_2,$ and $s_3$ are three absorbing states with different
  atomic propositions, so they are not (weak) bisimilar, as a result
  $s_{\mu_1},s_{\mu_2}$ and $s_{\mu_3}$ are not (weak) bisimilar
  either since they can evolve into $s_1,s_2,$ and $s_3$ with
  different probabilities. Therefore $s$ and $r$ are not (weak)
  bisimilar. Let $\phi=\MC{P}_{\geq0.4}(\X L(s_1))\land
  \MC{P}_{\geq0.3}(\X L(s_2))\land \MC{P}_{\geq0.3}(\X L(s_3))$, it is
  not hard to see that $s_{\mu_2}\models\phi$ but
  $s_{\mu_1},s_{\mu_3}\not\models\phi$, so $s\models \MC{P}_{\leq
    0}(\X\phi)$ while $r\not\models\MC{P}_{\leq 0}(\X\phi)$. When
  working in the setting of $\PA$s, $s_{\mu_1}$,
  $s_{\mu_2}$, and $s_{\mu_3}$ will not be considered as states, so we
  cannot use the above arguments for alternating automata any more.
\end{exa}

In the definition of $\iBS{1}$ and $\iSi{1}$, we choose first the
downward closed set $C$ before the successor distributions to be
matched, which is the key for achieving our new notions of
bisimulations and simulations. This approach was first adopted
in~\cite{AlfaroMRS07} to define the \emph{a priori metric} for Markov
decision processes, where it was shown that the a priori metric can be
characterized by the quantitative $\mu$-calculus. In~\cite{qapl} this
approach was also used to define a priori $\epsilon$-bisimulation
and simulation relations. 

\section{Conclusion and future work}\label{sec:conclusion}
In this paper we have introduced novel notions of bisimulation for $\PA$s. They are coarser than the existing bisimulations, and most importantly, we show that they agree with the logical equivalences induced by $\PCTL^{*}$ and its sub logics.
Even though we have not considered actions, it is
worth noting that actions can be easily added, and all the (weak) bisimulations
can be defined directly.
On the other hand, the (weak) bisimulations are then strictly finer than the logical equivalences,
because of the presence of these actions, similarly for simulations.

As future work, we plan to study decision algorithms for our new
(strong and weak) bisimulation and simulation relations.

\section*{Acknowledgement}
The authors are supported by IDEA4CPS and the VKR Center of Excellence  MT-LAB. The work has received support from the EU FP7-ICT projects TREsPASS (318003) and MEALS (295261), and the DFG Sonderforschungsbereich AVACS. Part of the work was done while the first author was with IT University of Copenhagen, Denmark. We thank Johann Schuster for detailed comments on an early version of this draft.

\bibliographystyle{abbrv}
\bibliography{bib}

\end{document}